%% file: main.tex
\documentclass[sigconf]{acmart}
\setcopyright{none}
\renewcommand\footnotetextcopyrightpermission[1]{}
\usepackage{hyperref}
\usepackage{listings}
\usepackage{booktabs}
\usepackage{url}
\usepackage{graphicx}
\usepackage{array}
\usepackage{soul}
\usepackage{color}
\usepackage{listings}
\usepackage{xspace}
\usepackage{multirow}
\usepackage{subfigure}

\makeatletter
\newcommand{\linebreakand}{%
  \end{@IEEEauthorhalign}
  \hfill\mbox{}\par
  \mbox{}\hfill\begin{@IEEEauthorhalign}
}
\makeatother

\newcounter{observation}
\newcommand{\observation}[1]{\refstepcounter{observation}
	\begin{center}
		\framebox{
			\begin{minipage}{0.93\columnwidth}
				{\bf Observation \arabic{observation}:} \textit{#1}
			\end{minipage}
		}
	\end{center}
}

\newcommand{\toolname}{\textit{PythonChangeMiner}\xspace}

\newcolumntype{L}[1]{>{\raggedright\let\newline\\\arraybackslash\hspace{0pt}}m{#1}}
\newcolumntype{C}[1]{>{\centering\let\newline\\\arraybackslash\hspace{0pt}}m{#1}}
\newcolumntype{R}[1]{>{\raggedleft\let\newline\\\arraybackslash\hspace{0pt}}m{#1}}

\begin{document}

\title[Changes from the Trenches: Should We Automate Them?]{Changes from the Trenches:\\ Should We Automate Them?}

\author{Yaroslav Golubev}
\affiliation{
    \institution{JetBrains Research}
    \city{Saint Petersburg}
    \country{Russia}
}
\email{yaroslav.golubev@jetbrains.com}

\author{Jiawei Li}
\affiliation{
    \institution{University of California, Irvine}
    \city{Irvine, CA}
    \country{United States}
}
\email{jiawl28@uci.edu}

\author{Viacheslav Bushev}
\affiliation{
    \institution{Saint Petersburg State University}
    \city{Saint Petersburg}
    \country{Russia}
}
\email{stardust.skg@gmail.com}

\author{Timofey Bryksin}
\affiliation{
    \institution{JetBrains Research}
    \institution{Higher School of Economics}
    \city{Saint Petersburg}
    \country{Russia}
}
\email{timofey.bryksin@jetbrains.com}

\author{Iftekhar Ahmed}
\affiliation{
    \institution{University of California, Irvine}
    \city{Irvine, CA}
    \country{United States}
}
\email{iftekha@uci.edu}

\begin{abstract}

Code changes constitute one of the most important features of software evolution. Studying them can provide insights into the nature of software development and also lead to practical solutions --- recommendations and automations of popular changes for developers.

In our work, we developed a tool called \toolname that allows to discover code change patterns in the histories of Python projects. We validated the tool and then employed it to discover patterns in the dataset of 120 projects from four different domains of software engineering. We manually categorized patterns that occur in more than one project from the standpoint of their structure and content, and compared different domains and patterns in that regard. We conducted a survey of the authors of the discovered changes: 82.9\% of them said that they can give the change a name and 57.9\% expressed their desire to have the changes automated, indicating the ability of the tool to discover valuable patterns. Finally, we interviewed 9 members of a popular integrated development environment (IDE) development team to estimate the feasibility of automating the discovered changes. It was revealed that independence from the context and high precision made a pattern a better candidate for automation. The patterns received mainly positive reviews and several were ranked as very likely for automation.

\end{abstract}

\maketitle

\input{sections/01-introduction}
\input{sections/02-background}
\input{sections/03-tool}
\input{sections/04-methodology}
\input{sections/05-results}
\input{sections/06-discussion}
\input{sections/07-threats}
\input{sections/08-conclusion}

\bibliographystyle{ACM-Reference-Format}
\bibliography{icse}

\end{document}

%% file: sections/01-introduction.tex
\section{Introduction}

Software engineering studies encapsulate all steps of the software development life-cycle~\cite{rani2017detailed}, ranging from requirement analysis~\cite{chakraborty2012role} to testing~\cite{garousi2017worlds}, deployment~\cite{rodriguez2017continuous}, and maintenance~\cite{lenarduzzi2017analyzing} of the software. Naturally, all of them include \textit{changes}, which can happen for various reasons, such as refactoring~\cite{palomba2017exploratory}, bug fixes~\cite{cotroneo2019analyzing}, implementation of new features, etc.

These changes can be unique; however, in many cases, they are repetitive and follow \textit{patterns}~\cite{nguyen2013study}. Such patterns can be a rich source of information for analyzing the history of changes and their impact~\cite{kawrykow2011non}, modification records of fault fixes~\cite{german2006empirical}, or code change patterns' relationship with adaptive maintenance~\cite{meqdadi2020study}.

Another important use of such code change patterns are various \textit{suggestions} and \textit{automations}, i.e. mining popular changes from the existing code and suggesting them to developers or applying them automatically~\cite{ying2004predicting, dagenais2011recommending, nguyen2016api}. To date, most of the existing work in this research area focus on statically-typed languages such as Java. However, Python has been gaining popularity, especially in the areas of machine learning and data analysis~\cite{nagpal2019python}, and is significantly less studied. This, in turn, impedes researchers from answering fundamental questions, such as: what are the most frequent changes done by developers in Python, how many of these changes follow a pattern, can they be automated, etc.
Our goal in this study is to fill this gap in understanding by answering the aforementioned questions, which also has actionable practical implications in the form of discovering potential candidates for automations.

\begin{figure}[t]
\centering
\subfigure[Commit in the Cupy project~\cite{cupy_commit}.]{
	\label{subfig:correct}
	\includegraphics[height=0.27in]{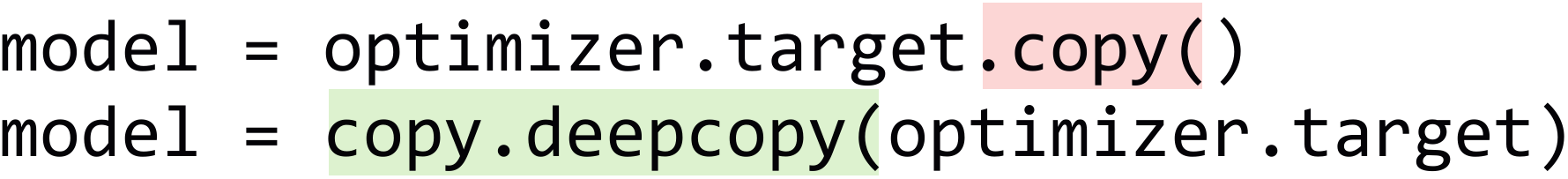}} 
	\hfill
\subfigure[Commit in the Ray project~\cite{ray_commit}.]{
	\label{subfig:notwhitelight}
	\includegraphics[height=0.27in]{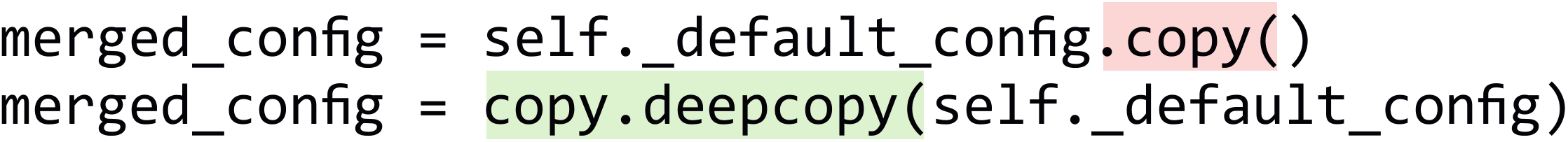}} 
	\hfill
\subfigure[Commit in the SciKit Multiflow project~\cite{scikit_commit}.]{
	\label{subfig:nonkohler}
	\includegraphics[height=0.27in]{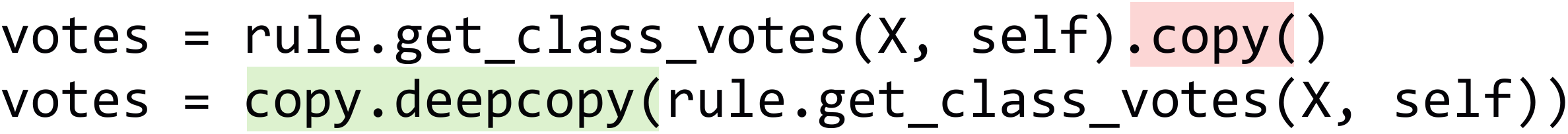}} 
	\vspace{-0.3cm}
\caption{An example of a change pattern identified in several projects on GitHub. The developers switched from using built-in copying to creating a deep copy of an object using a \texttt{copy} module of a standard library.}
\vspace{-0.3cm}
\label{fig:example_pattern}
\end{figure}

Since our goal is to conduct a large-scale analysis of code changes and to answer the questions mentioned above, a complete manual analysis is not feasible. Therefore, we start by developing a tool to make the process automated. We developed a tool called \toolname for mining change patterns in Python code using a special code representation called fine-grained program dependence graphs (fgPDGs) that was introduced for Java by Nguyen et al.~\cite{nguyen2019graph}. The tool analyzes the version control system history of a project, parses each commit into graphs, and then uses a recursive algorithm to detect repetitive patterns within the changes. We validated \toolname on a small dataset by manually checking whether the discovered code patterns represent actual repeated changes. The tool demonstrates a 97.2\% precision. An example change pattern that appears in several projects is presented in Figure~\ref{fig:example_pattern}.

To gain an understanding regarding the most frequent changes in Python, we conducted a large-scale mining analysis. We analyzed 120 projects collected from four \textit{domains}: Web, Media, Data processing, and Machine Learning + Deep Learning. To keep the operation time of the search reasonable, in this study we focused on changes that involve function calls. We discovered a total of 7,481 patterns, of which 803 appeared in at least two projects. To better understand the changes, we investigated the recurring patterns in detail and classified them in two different dimensions: \textit{structurally} and \textit{thematically}. This allowed us to draw some observations and compare the domains between each other. To evaluate the meaningfulness of the patterns, as well as to discover whether the mined patterns can serve as a source of potential automations for integrated development environments (IDEs), we surveyed the authors of the code changes in the patterns. 82.9\% of the respondents were able to give their change a name and 57.9\% affirmed that they would like to have the change automated for them in the IDE.

Finally, to get the IDE developers' perspective regarding the patterns, we interviewed the members of the development team of PyCharm,\footnote{PyCharm, an IDE for Python: \url{https://www.jetbrains.com/pycharm/}} a popular IDE for Python. We asked them to rate 15 patterns based on their potential usefulness and perceived difficulty in automating them. 

Overall, our contributions are:

\begin{itemize}
    \item Developed and validated a tool called \toolname that mines change patterns in Python code using fine-grained program dependence graphs.
    \item Conducted a study of code change patterns from 120 projects in 4 domains, categorized recurring patterns based on their structure and topic to gather insights about the nature of changes.
    \item Conducted a survey of the authors of the changes about the discovered change patterns, with 57.9\% of the respondents indicating that they want to have their prior change automated.
    \item Interviewed 9 members from the PyCharm development team, compared their rankings of patterns with those of the developers from the survey, and discovered change patterns with the highest potential to be turned into a suggestion for the user.
\end{itemize}

The remainder of the paper is organized as follows. In Section~\ref{sec:background}, we mention existing work in mining code changes and empirical studies of coding practices and automation. Section~\ref{sec:tool} describes the tool that we developed and its validation, and Section~\ref{sec:method} describes the methodology of our empirical study. In sections~\ref{sec:result} and~\ref{sec:discussion} we talk about the results of our study and their implications, in Section~\ref{sec:threats} we list possible threats to validity of our research, and in Section~\ref{sec:conclusions} we draw up conclusions.

%% file: sections/02-background.tex
\section{Related Work \& Background}
\label{sec:background}

\subsection{Code Changes}
A lot of work has been dedicated to investigating changes in software projects to get a deeper insight into the nature of software development. Ying et al.~\cite{ying2004predicting} developed an approach that applies frequent pattern mining to determine code change patterns from code change history. Zimmermann et al.~\cite{zimmermann2005mining} applied another set of data mining techniques to a code version history to explore intricacies in code changes, then based on this information they evaluated its effectiveness for predicting further code changes. More recently, Nguyen et al.~\cite{nguyen2013study} conducted a large-scale study of the repetitiveness of code changes in software evolution. The authors found that the repetitiveness of code changes decreases as change sizes increase, and also that bug-fixing changes repeat similarly to general ones, which can be beneficial for recommendation systems and automated program repair. Other works explored the characteristics of bug-fixing change patterns~\cite{osman2014mining, martinez2014accurate, zhao2017towards, koyuncu2020fixminer} and prediction of faults based on changes~\cite{hassan2009predicting, kim2006automatic, kim2008classifying, mockus2000predicting, sliwerski2005changes, shivaji2012reducing}.  Negara et al.~\cite{negara2014mining} designed an algorithm to mine previously unknown code change patterns from fine-grained sequence of code changes recorded from IDEs, identified ten interesting change patterns, and conducted a survey regarding the relationship between those patterns and developers' activities. They also analyzed the developers' preference for automating these changes in IDEs. 

A large number of these works have studied traditional statically-typed languages such as C/C++ and Java. However, dynamic languages were also considered. Lin et al.~\cite{lin2016empirical} implemented an automatic tool, \textit{PyCT}, to extract multiple types of fine-grained Python source code changes from commit history information of ten Python projects across five domains, with the goal of investigating the characteristics of Python source code changes and finding insights on software evolution. Chen et al.~\cite{chen2018study} analyzed the relation between bug-fixing activities and fine-grained changes of dynamic feature code in seventeen Python projects. They provided valuable information on how developers handle changes of dynamic feature code when fixing bugs. Controneo et al.~\cite{cotroneo2019analyzing} conducted an empirical study on three Python projects and found that the locations of recurring bug-fixing change patterns are specific source code contexts.

In our work, we aim to combine different aspects of the mentioned works for Python. To the best of our knowledge, no work before has been directed specifically towards mining a large number of code change patterns from Python code and then studying them from the standpoint of searching for potential automations in an IDE.

\vspace{-0.3cm}

\subsection{CPatMiner}
\label{sec:cpatminer}
Nguyen et al.~\cite{nguyen2019graph} presented an algorithm for mining previously unknown semantic patterns in Java code called \textit{CPatMiner}. The logic of the authors is that relying only on syntactic changes may lead to wrongful definitions of patterns: performing identical changes to the abstract syntax tree (AST) does not always indicate the semantic closeness of the changes. 

To mitigate this limitation, the authors rely on the representation of code that they call \textit{fine-grained program dependence graph (fgPDG)}. This graph includes three types of nodes: \textit{data} nodes (variables, literals, constants, etc.), \textit{operation} nodes (arithmetic, bit-wise operations, etc.), and \textit{control} nodes (control sequences like \textit{if}, \textit{while}, \textit{for}, etc.). These nodes can be linked by two types of edges: \textit{control} edges represent a connection between a \textit{control} node and a node that it controls, and \textit{data} edges show the flow of the data in the program, such edges also have labels specifying the flow of data.

The authors then employ this representation towards code changes by mapping fgPDGs of the versions of code before and after the change, resulting in \textit{change graphs}. Corresponding nodes from the two versions are connected by special \textit{map} edges, and the authors define semantic change patterns as such change graphs that are repeated several times within the dataset, with the threshold defined by the user. The authors implement the algorithm for searching these patterns in a given dataset of projects and demonstrate that the patterns they discover represent real repeated changes using a survey of developers.

With a growing popularity of Python, especially in relevant and developing areas such as data analysis, machine learning, and deep learning, it is of great interest to discover similar  patterns for Python. However, the parser in \textit{CPatMiner} is written specifically for the syntax of the Java language, and the tool stores graphs and works with them as Java objects, so we cannot reuse the tool directly. At the same time, the algorithm itself is not language-specific, because it relies only on the AST of code before and after the change.

For that reason, to discover semantic change patterns in Python, we used CPatMiner's algorithm to implement our own tool called \toolname\footnote{\toolname: \url{https://github.com/JetBrains-Research/python-change-miner}} aimed specifically at Python. We demonstrate its capabilities by conducting an empirical study on a large dataset of projects, conducting a survey of GitHub developers, and gathering insights from IDE development team members.

%% file: sections/03-tool.tex
\section{PythonChangeMiner}
\label{sec:tool}

\subsection{Tool Description}

The tool's operation logic follows that of \textit{CPatMiner} closely, as described in Section~\ref{sec:cpatminer}. In this section, we provide a brief description of how \toolname works. A detailed description of the algorithm is available in the \textit{CPatMiner} paper~\cite{nguyen2019graph}.

\toolname can do the following operations on Python source files:

\begin{itemize}
    \item build fgPDGs for files;
    \item create change graphs for functions in two revisions of files;
    \item mine change graphs from the version control system history of a given Git project;
    \item mine patterns in the obtained change graphs.
\end{itemize}

The tool mines a history using the PyDriller framework~\cite{spadini2018pydriller} and builds change graphs for matching functions in each changed file of each commit. To do that, both versions of the file (before and after the change) are parsed into ASTs, which are traversed to create a fine-grained Program Dependence Graph described in Section~\ref{sec:cpatminer}. The tool uses GumTree~\cite{DBLP:conf/kbse/FalleriMBMM14} to identify the corresponding and changed nodes and then constructs a change graph from fgPDGs. 

\begin{figure}[t]
\centering
\includegraphics[width=3.3in]{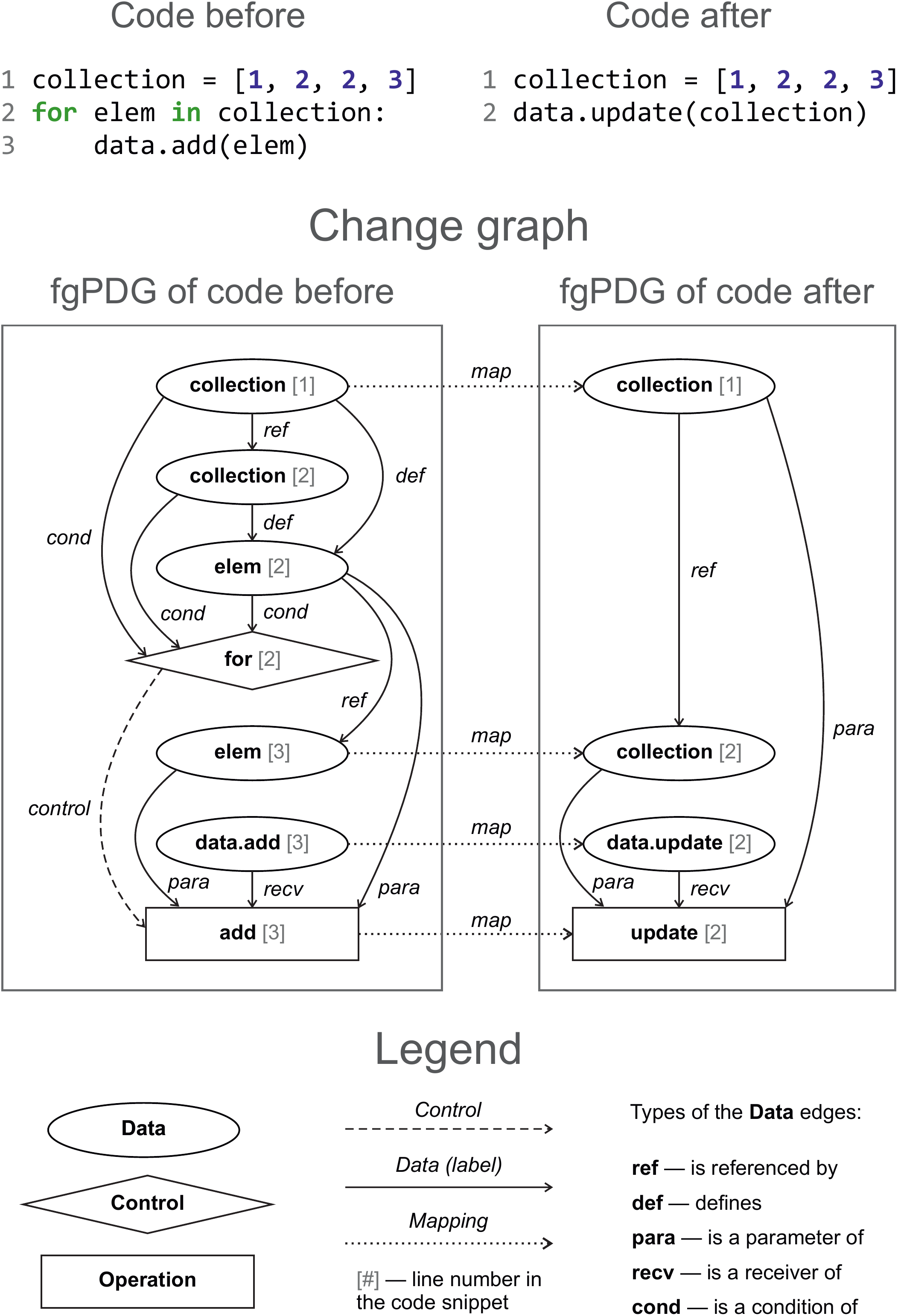}
\caption{An example of a code change and a corresponding code change graph.}
\vspace{-0.4cm}
\label{fig:cg}
\end{figure}

A code change and a corresponding change graph are presented in Figure~\ref{fig:cg}. This change shows how a set \texttt{data} is updated with the members of the list \texttt{collection}. The left and the right parts of the change graph are fgPDGs of the code before and after the change. They include \textit{data}, \textit{control}, and \textit{operation} nodes that are connected with \textit{data} and \textit{control} edges. \textit{Data} edges have additional labels that specify the type of the data flow, like reference, definition, or receiving as a parameter. For example, the fgPDG of the code before in Figure~\ref{fig:cg} introduces the variable \texttt{collection} and its reference, describes that variables \texttt{collection} and \texttt{elem} are parts of the \texttt{for} loop condition, and shows that the \texttt{for} loop controls the call of the \texttt{add} method.

The two fgPDGs are connected together with \textit{map} edges that join the corresponding nodes, creating a single united change graph. In our graph, these edges connect the same \texttt{collection} variables that did not change, as well as different objects in the last lines of the code snippets: the method, its parameter, and the object that calls it.

When all the necessary change graphs are built, they are analyzed to find all mapped node pairs before and after the change. To make the operation time of the tool reasonable and to lower the computational difficulty, only function call nodes are considered during this initial listing. All pairs of function call nodes connected by a \textit{map} edge are considered to be initial, trivial patterns. An example of such a pair is the bottom mapping between nodes \texttt{add} and \texttt{update} in Figure~\ref{fig:cg}. These patterns are then recursively extended to new nodes, and the user can define what patterns to detect by setting the minimum number of nodes in a pattern and its minimum frequency in the corpus. For the example in Figure~\ref{fig:cg}, if there are several cases in the dataset where \texttt{set.add(elem)} is changed to \texttt{set.update(list)}, then the pattern will grow to include the corresponding nodes, and if it is larger and more frequent than the given threshold, the pattern graph and the code of its samples will be in the output. From here on out, a \textit{pattern} means a repeated change, and a \textit{sample} of a pattern means a single specific instance of this pattern. After the search is finished, the tool saves the samples of patterns as graph files, a PDF visualization, and HTMLs of changed code that can be viewed and evaluated by researchers.

\subsection{Validation}
\label{sec:validation}

To validate the performance of the tool, we manually evaluated a number of patterns. In this validation, we consider a pattern to be \textit{correct} if it represents an actual repeated change, and \textit{incorrect} if it does not.

To obtain the patterns, we downloaded eight small projects from the GitHub \textit{Trending} page that fulfill three conditions: main language Python, no less than 100 stars, no more than 1,500 commits, following the guidelines in literature~\cite{kalliamvakou2014promises}. The list of projects is presented in Table~\ref{table:validation_projects}.

\begin{table}[h]
\centering
\vspace{-0.2cm}
\caption{The summary of projects used for the validation of \toolname. Contrib. stands for contributors, the age is in years and months, the dataset was collected in July, 2020.}
\vspace{-0.2cm}
\label{table:validation_projects}
\begin{tabular}{ c c c c }
\toprule
\textbf{Project} & \textbf{Commits} & \textbf{Contrib.} & \textbf{Age}\\
\midrule
trailofbits/algo & 1040 & 148 & 4 y. 2 m. \\
tiangolo/fastapi & 1126 & 180 & 1 y. 7 m. \\
unit8co/darts & 163 & 12 & 1 y. 10 m.\\
openai/gym & 1211 & 248 & 4 y. 3 m. \\
python-poetry/poetry & 1486 & 212 & 2 y. 5 m.\\
waditu/tushare & 412 & 12 & 5 y. 6 m.\\
elyra-ai/elyra & 535 & 12 & 2 y. 3 m.\\
open-mmlab/mmcv & 417 & 57 & 1 y. 11 m.\\
\bottomrule
\end{tabular}
\vspace{-0.2cm}
\end{table}

Since \toolname identifies changes of any size, we needed to filter out small changes to ensure that the identified change patterns are non-trivial. We conducted a preliminary manual analysis, where we tried to identify thresholds for filtering out small changes. We decided on the following thresholds for a pattern: the minimum number of nodes in the graph (the size of the pattern) --- 4, and the minimum number of samples in the corpus (the frequency of the pattern) --- 3. In total, 2,121 change graphs were extracted and processed, resulting in 72 patterns containing 369 samples. The authors of the paper independently reviewed all the patterns and then discussed the results, reaching a unanimous decision for each pattern. 

70 (97.2\%) of the inspected patterns were straightforward and actually represented change patterns occurring in different places.
Only 2 (2.8\%) patterns were deemed incorrect. In one pattern, a variable was mapped to the sum of a list and another variable, and in the other pattern, the samples were methods named \texttt{.startswith()}, and had nothing in common except for this name. We consider the obtained accuracy to be sufficient for the purposes of this study. It can be seen that \toolname provides a reliable outcome of correct patterns that contain function calls and can be used for mining them on GitHub.

%% file: sections/04-methodology.tex
\section{Methodology}
\label{sec:method}

The aim of our study lies in researching popular code change patterns in Python code, understanding such changes and, finally, identifying changes that can be implemented as suggestions in IDEs. We started by compiling a dataset of various Python projects from GitHub, searched them for patterns, categorized and analyzed the discovered patterns, surveyed their authors, and interviewed the members of the PyCharm development team whether the patterns were suitable for automation. 

Since we survey and interview two different groups of developers, and because the term \textit{developer} is very broad, we use the following names in order not to confuse the reader. We refer to the authors of the changes as \textbf{GitHub authors}, and describe the \textbf{GitHub authors survey}. On the other hand, we refer to the developers of the PyCharm as \textbf{IDE team members}, and describe the \textbf{IDE team interview}.

\begin{figure}[h]
\centering
\includegraphics[width=\columnwidth]{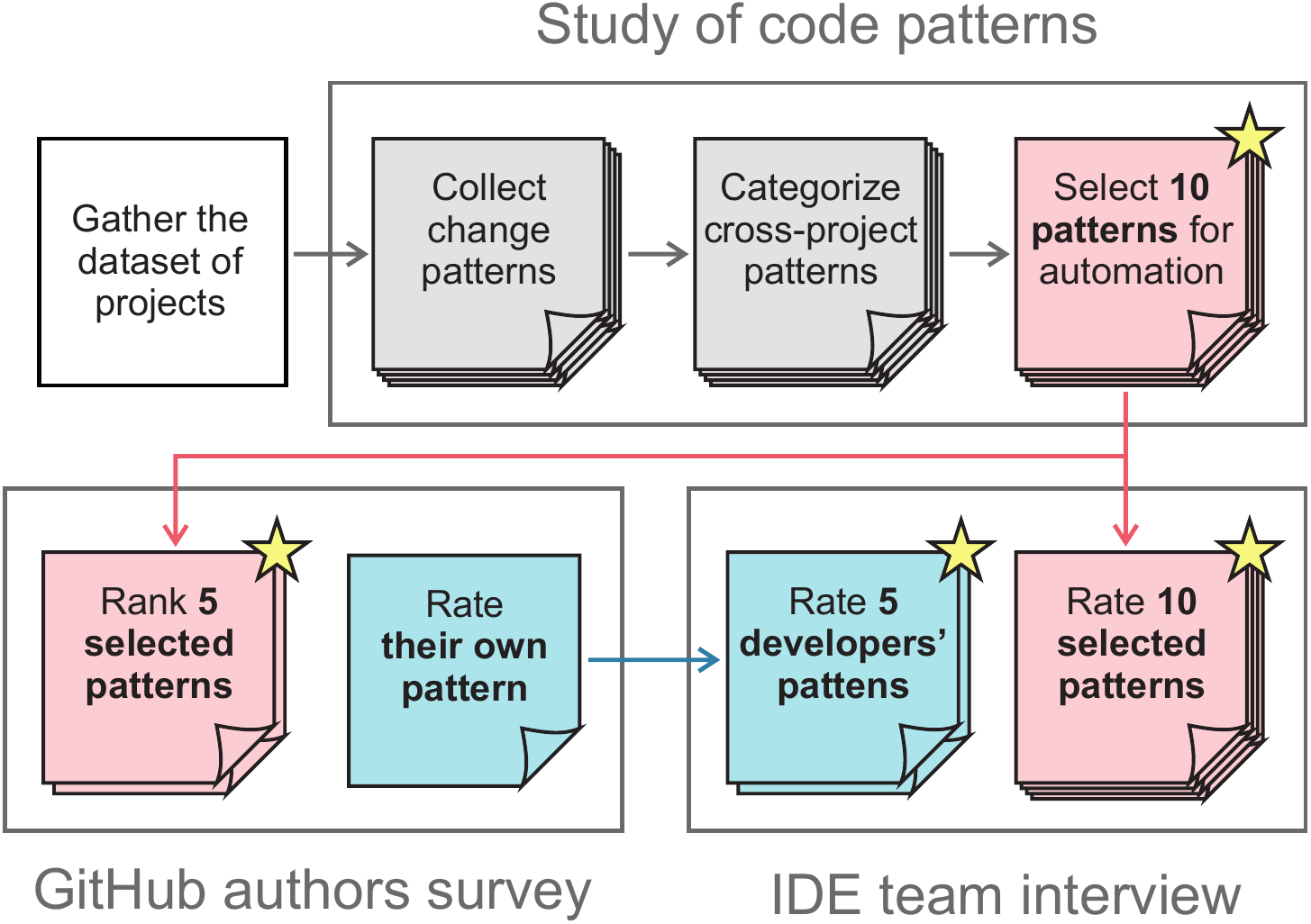}
\caption{The pipeline of the study. For the survey and the interview, the figure shows what each participant does. The star marks the patterns selected by the authors of this paper.}
\vspace{-0.5cm}
\label{fig:pipeline}
\end{figure}

The pipeline of the study is presented in Figure~\ref{fig:pipeline}, and the details of each step are described below.

\subsection{Dataset}
It is of interest not only to discover the most popular code change patterns in a large number of projects, but also to find patterns that exist across different projects. Therefore, we decided to compile our dataset with an equal number of projects from different domains.

Lin et al.~\cite{lin2016empirical} use the following domains in their study: \textit{Web}, \textit{Data processing}, \textit{Science computing}, \textit{NLP}, and \textit{Media}, which allows them to present an interesting comparison. In our study, we set out to gather a larger dataset of more projects, and also decided to extend the NLP domain to more diverse machine learning repositories. We ended up with the following four domains: \textbf{Web} (frameworks, tools for downloading, archiving), \textbf{Media} (graphics editors, video editors, music library managers), \textbf{Data} (libraries for mathematics and computations, tools for working with tables, plotting graphs, and other data processing), \textbf{ML+DL} (everything that has to do with machine learning, natural language processing, neural networks).

For each of these four domains, we picked 30 Python projects that meet the following four conditions, following the guidelines in literature~\cite{kalliamvakou2014promises}: not a clone of any other repository in the list, number of commits between 1,000 and 30,000, age of the project at least 2 years, and at least 10 contributors. The dataset was compiled in July, 2020, the full list of projects is available.\footnote{The dataset: \url{https://zenodo.org/record/4004118}}

We searched for patterns in the dataset using the same settings as during preliminary experiments: the minimum number of nodes in the graph (the size of the pattern) --- 4, and the minimum number of samples in the corpus (the frequency of the pattern) --- 3. 

\subsection{Categorization}
\label{sec:categorization}

Since the tool discovers a large number of patterns, we decided to limit our focus of manual study to the patterns that occurred in at least two different repositories (\textit{cross-project} patterns). The logic behind this is that such patterns should be more general by definition and, therefore, are more interesting from the standpoint of discovering possible suggestions, and other automations.

To better understand the nature of the discovered changes, we decided to categorize all the studied patterns in two different dimensions. Since all of the patterns have to do with functions and function calls, they can be classified \textit{structurally} (indicating where the function comes from and how it is applied) and \textit{thematically} (indicating what functionality the change relates to). These classifications can potentially allow us to gather insights about the use of the changes involving function calls and compare projects and domains.

To categorize the patterns, the first two authors of the paper worked together, using the open coding technique~\cite{sharma2015s} in three iterations. During the first iteration, we evaluated all changes and gave them short descriptions. During the second iteration, we clustered similar change patterns, separately for the structural and thematic categories. When disagreements arose during the second stage, the two authors resolved them with a discussion, with all the results being unanimous. Finally, during the third iteration, we analyzed the clusters and obtained the final short lists of categories.

Since our goal is to select candidate patterns for automation and due to a large number of cross-project patterns ($\approx$800), it is not feasible to validate all these patterns with practitioners. For this reason, we selected 100 most frequent patterns that show the best potential to be automated in an IDE. Then all the authors of the paper discussed these patterns until we were left with 10 final candidates. When selecting code change patterns that can be automated, the authors based their decision on their own experience of Python development and the diversity of patterns categories obtained during the open coding: the final 10 patterns selected to be shown to GitHub authors and IDE team members include patterns from almost all categories.

\subsection{GitHub Authors Survey}
\label{sec:methdology_gitsurvey}
To get a deeper insight into the nature of the discovered changes, we surveyed their authors. To do that, we collected the emails that were used as a signature for each git commit containing the change and sent an email to each author. Since the emails could repeat among different samples and commits, we sent a single email to each of the addresses. In total, there are 1,585 unique emails in our dataset. After filtering out emails ending with \textit{noreply.github.com}, incorrect emails, and receiving a lot of them undelivered, we successfully sent 1,364 emails. In total, we received 76 answers resulting in a response rate of 5.5\%, which we considered satisfactory, as it is close to other studies in the field that had reported a response rate between 5.7\%~\cite{PassosTSE} and 7.9\%~\cite{FlavioMedeiros_2018}.

During the survey, the participants were shown their own change and asked to fill in a survey with the following questions: 

\begin{itemize}
    \item How many years of software development experience do you have?
    \item Briefly describe the highlighted change and why you made it, if possible.
    \item Can you give it a name? If yes, please propose a name for the change. If no, please comment on why.
    \item Would you like to have this change automated by a tool (no matter how difficult that automation might be)? If no, why?
\end{itemize}

Additionally, we randomly picked 5 out of 10 selected patterns (see Section~\ref{sec:categorization}) in order not to overload the respondents and showed them to each of the participants (the same 5 for each) alongside their own change. The respondents were asked to rank the changes from the most useful in terms of automation to the least useful based on their perception. We used a separate ranking for this question.

\subsection{IDE Team Interview}
\label{sec:methdology_ide}

To evaluate the feasibility of automating the discovered changes, we got a second opinion from the developers of an IDE. We interviewed 9 members of the development team of PyCharm, a popular IDE for Python.

The PyCharm team members were shown representative samples from 15 change patterns in random order. 10 of them are the selected patterns described above (see Section~\ref{sec:categorization}). The other 5 were selected in the following fashion. We gathered all the patterns that their authors in the survey said they would like to have automated (described in Section~\ref{sec:methdology_gitsurvey}). Then, we once again applied open coding, with the same methodology as described in Section~\ref{sec:categorization} to classify the changes by the feasibility of their automation. The idea behind this classification is to see what people want automated and estimate why some of these changes are not suitable for general automations. Finally, we picked the necessary 5 patterns from those that do not have any specific limitations for automation.

For each of 15 patterns, we provided a small description of its purpose. We used the original commit messages to write these descriptions. We also wanted our questions to PyCharm team members to be more specific than those for the authors of the changes, because IDE developers can provide insights into how realistic the idea of automating the change is. For that reason, we asked them specifically about turning each change into an \textit{inspection} or an \textit{intention}. These are terms related to the IntelliJ platform,\footnote{IntelliJ: \url{https://jetbrains.org/intellij/sdk/docs/intro/intellij_platform.html}} which PyCharm, IntelliJ IDEA, CLion, and other JetBrains IDEs are built upon: the code is being analyzed while you write it, and the IDE provides inspections for locating and fixing anomalous or bad code (highlighting the code) or intentions for when you can optimize the code (a yellow lightbulb near the code). Thus, we asked the PyCharm team the following three questions:

\begin{itemize}
    \item Do you think that this change should be suggested to the user via an inspection or an intention in PyCharm?
    \item How difficult do you think it would be to turn this change into an inspection or an intention? 
    \item Why do you think that this change should or should not be automated?
\end{itemize}

This allows us not only to find patterns with most potential, but also to compare the opinions of GitHub authors and the IDE team members with each other.

%% file: sections/05-results.tex
\section{Results}
\label{sec:result}

\subsection{Change Patterns}

Running \toolname on our dataset resulted in discovering 7,481 patterns that altogether contain 37,623 samples of code. The smallest patterns have 3 samples (according to the threshold we used to define a pattern), the largest pattern has 156 samples (all of them in a single commit of a single repository). Of these 7,481 patterns, 803 include samples from at least two different projects.

In regards to the four domains, 243 patterns include at least one sample from Web, 384 from Media, 399 from Data, and 365 from ML+DL. Media has the largest number of patterns that include only samples from one domain. Through manual analysis we observed that the main reason for this includes relying on the same external libraries and functions such as \textit{GTK}. Working with UI, drawing objects, etc. requires using specific APIs. On the other hand, patterns that occur within all four domains contain the most general functionality. An example of such a pattern is changing the condition from \texttt{os.path.exists()} to \texttt{os.path.isfile()}, thus specifying it more carefully. Other changes here deal with loggers, file paths, error types, and assert conditions in  tests.

\subsection{Categorization}

After applying open coding as described in Section~\ref{sec:categorization}, we obtained 5 structural categories of change patterns and 9 thematic categories (8 + Other). In this section, we describe the categories and their distribution in more detail.

\subsubsection{Structural Categories}

Structurally, the patterns can be divided as follows:

\begin{itemize}
    \item \textbf{Built-in Functions} include patterns where the change has to do with Python's built-in, default functions and methods, like \texttt{str()} or \texttt{.upper()}.
    \item \textbf{Standard Library} patterns include changes to functions from the standard library,\footnote{Python Standard Library: \url{https://docs.python.org/3/library/}} either within the same function or consisting in the change of function. The difference from the \textbf{Built-in Functions} is that using functions from the standard library requires importing them first.
    \item \textbf{External Library} patterns are similar to the previous category, but relate to external libraries that have to be installed.
    \item \textbf{Original Functions} include patterns that cover changes to original functions written by the developers of the project.
    \item \textbf{Moved Functionality} occurs when the change shifts the functionality between the categories described above. This can mean changing the library, moving from a built-in solution to a library, or even moving from a library to a self-defined function.
\end{itemize}

\begin{figure}[h]
\centering
\includegraphics[width=3.0in]{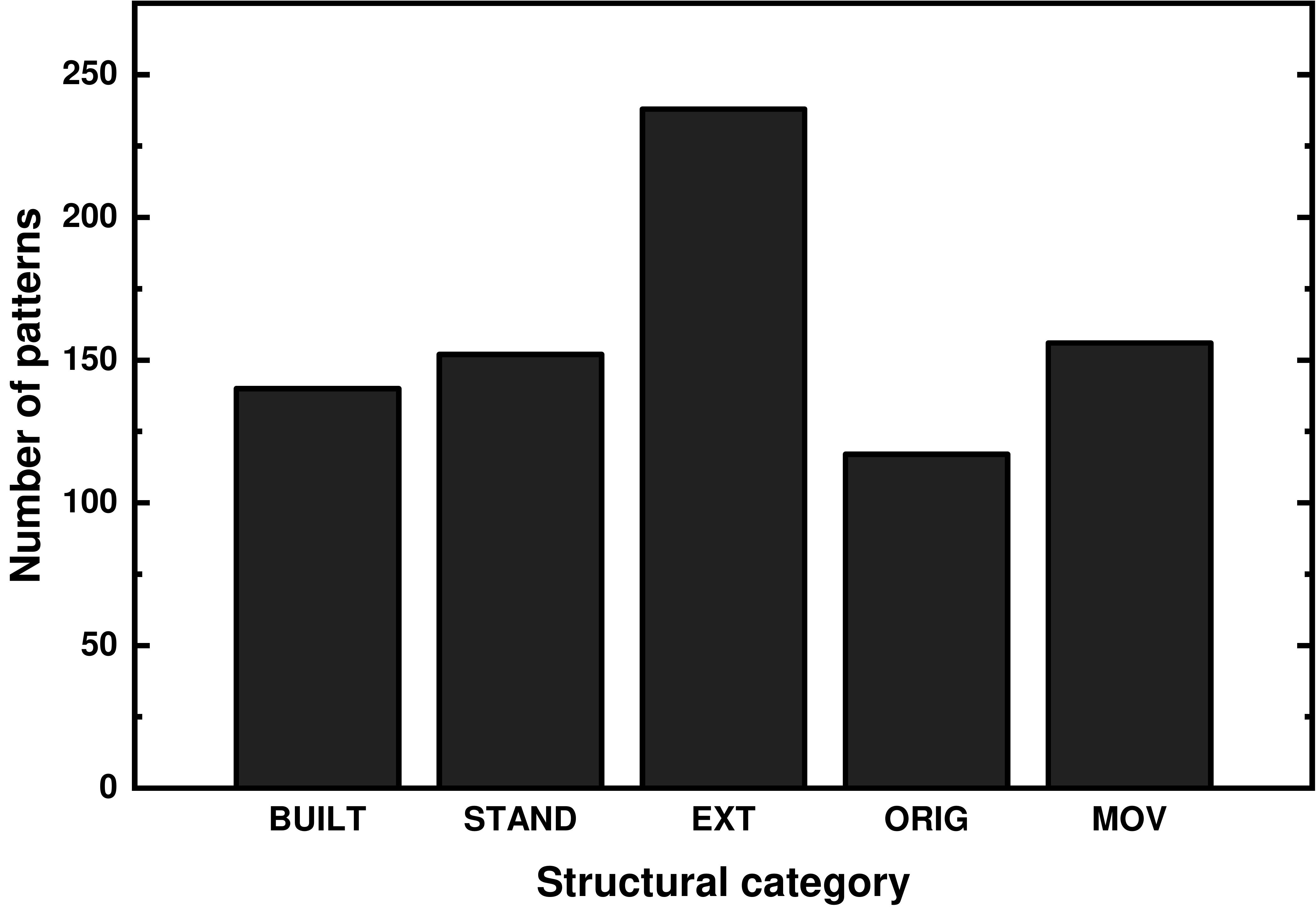}
\vspace{-0.1cm}
\caption{The distribution of patterns by their structure. \textit{BUILT} stands for \textit{Built-in Functions}, \textit{STAND} -- \textit{Standard Library}, \textit{EXT} -- \textit{External Library}, \textit{ORIG} -- \textit{Original Functions}, \textit{MOV} -- \textit{Moved Functionality}.}
\label{fig:structural}
\vspace{-0.1cm}
\end{figure}

Such a classification allows us to understand where the changed functions come from. \textit{Moved Functionality} is especially interesting, because such changes might indicate dissatisfaction with some original functionality. 
Figure~\ref{fig:structural} shows the general distribution of patterns by their structural categories.

\observation{Half of all the discovered patterns describe changes to functions within a single library.}

This half is comprised of the Standard Library patterns and External Library patterns, where in both cases the change occurs within a single library. It can be also seen that changes to functions in the External Library category are more frequent than changes to functions in the Standard Library. The external libraries that cover most changes in our dataset are \texttt{numpy}, \texttt{tensorflow}, \texttt{pandas}, \texttt{pytorch}, and there is a long list of other smaller libraries. The most popular modules of the standard library include \texttt{logging}, \texttt{unittest}, \texttt{os}, \texttt{re}.

The rest of the categories are spread out more or less equally. From the standpoint of gathering insights, the most interesting category is Moved Functionality. 

\observation{Changes that move from built-in functions to libraries are often made to add new functionality or simplify the code.}

For example, one popular pattern consists of migrating the code from \texttt{object.copy()} to \texttt{copy.deepcopy(object)}, thus making a deep copy of the object, as demonstrated in Figure~\ref{fig:example_pattern} above. This change was encountered in several ML+DL repositories. One commit message indicated that the copying was related to a state of a machine learning model. Another example that includes moving to an external library is moving from \texttt{set()} to \texttt{numpy.unique()}, which returns unique objects as a sorted array.

Sometimes, the change happens between different libraries. A common change is switching from \texttt{os.rename} to \texttt{shutil.move}. One commit message explained that \texttt{shutil.move} actually allows to move files between different file systems (for example, to mobile devices, which is necessary in a lot of user-based applications). On the other hand, a less frequent pattern but still occurring in several projects is the opposite change. The commit message specifies that within one file system \texttt{os.rename} provides a more stable result.

\begin{figure}[h]
\centering
\vspace{-0.2cm}
\includegraphics[width=3.0in]{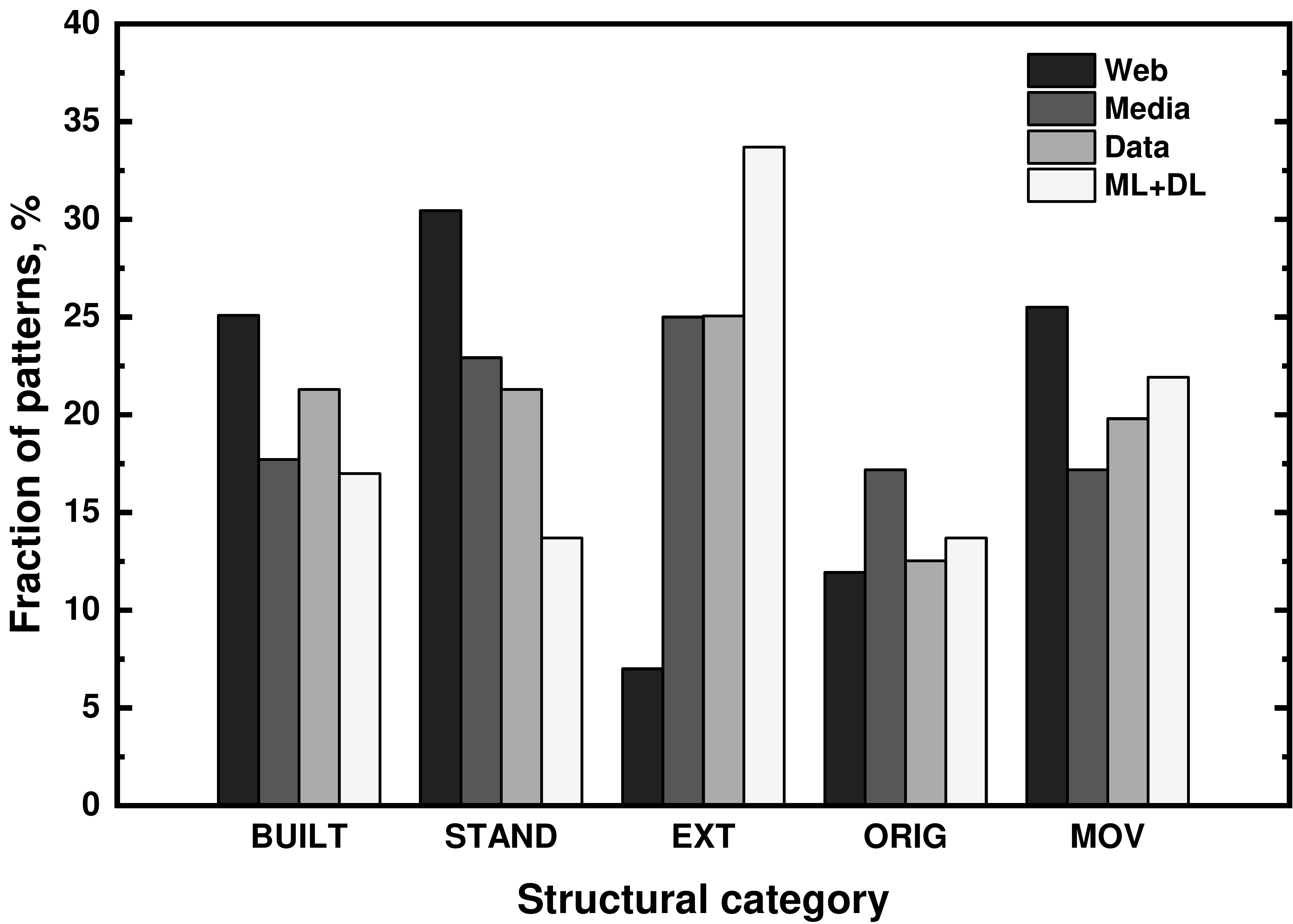}
\vspace{-0.2cm}
\caption{The distribution of patterns by their structure in different domains. \textit{BUILT} stands for \textit{Built-in Functions}, \textit{STAND} -- \textit{Standard Library}, \textit{EXT} -- \textit{External Library}, \textit{ORIG} -- \textit{Original Functions}, \textit{MOV} -- \textit{Moved Functionality}.}
\label{fig:structural_domains}
\vspace{-0.4cm}
\end{figure}

It may be also interesting to see how the structure differs between projects in different domains. Here we consider a pattern to be included in the domain if it occurs in at least one project from this domain. Figure~\ref{fig:structural_domains} shows the same distribution by structural categories as Figure~\ref{fig:structural}, but with specific domains.

\observation{In Web, significantly fewer changes relate to functions from external libraries.}

We suspect that this happens because every other domain except Web heavily relies on specific external libraries: Media relies on GTK and other UI-related libraries, while Data and ML+DL rely on \texttt{numpy}, \texttt{pandas}, \texttt{tensorflow}, and \texttt{pytorch}. Because Web does not have such reliance, it changes such functions less often compared to other categories. 

\subsubsection{Thematic categories}

Thematically, open coding resulted in the following categories:

\begin{itemize}
    \item \textbf{Data Structures} involve changes to the data containers or their initialization. This can be switching from a list to a set or initializing a matrix differently.
    \item \textbf{Data Processing} involves various manipulation with data. This includes copying objects, working with classes and their parameters, reshaping arrays, etc.
    \item \textbf{Calculations} include everything that has to with mathematics and numbers.
    \item \textbf{Conditions} include everything boolean, like asserts and checks.
    \item \textbf{Text} includes working with strings, like regular expressions, encoding, decoding, etc.
    \item \textbf{I/O} includes everything that the user sees, like logging, printing, errors, or arguments parsing.
    \item \textbf{Files} cover working with the file system, saving, and reading files.
    \item \textbf{None} relates to changes that change absolutely nothing from the meaningful side of code, for example, a refactoring that moves from importing the entire library and then using a function from it to importing only this function and then using its full name.
    \item Finally, \textbf{Other} covers everything that does not quite fall into the categories described above.
\end{itemize}

\begin{figure}[h]
\centering
\vspace{-0.2cm}
\includegraphics[width=3.0in]{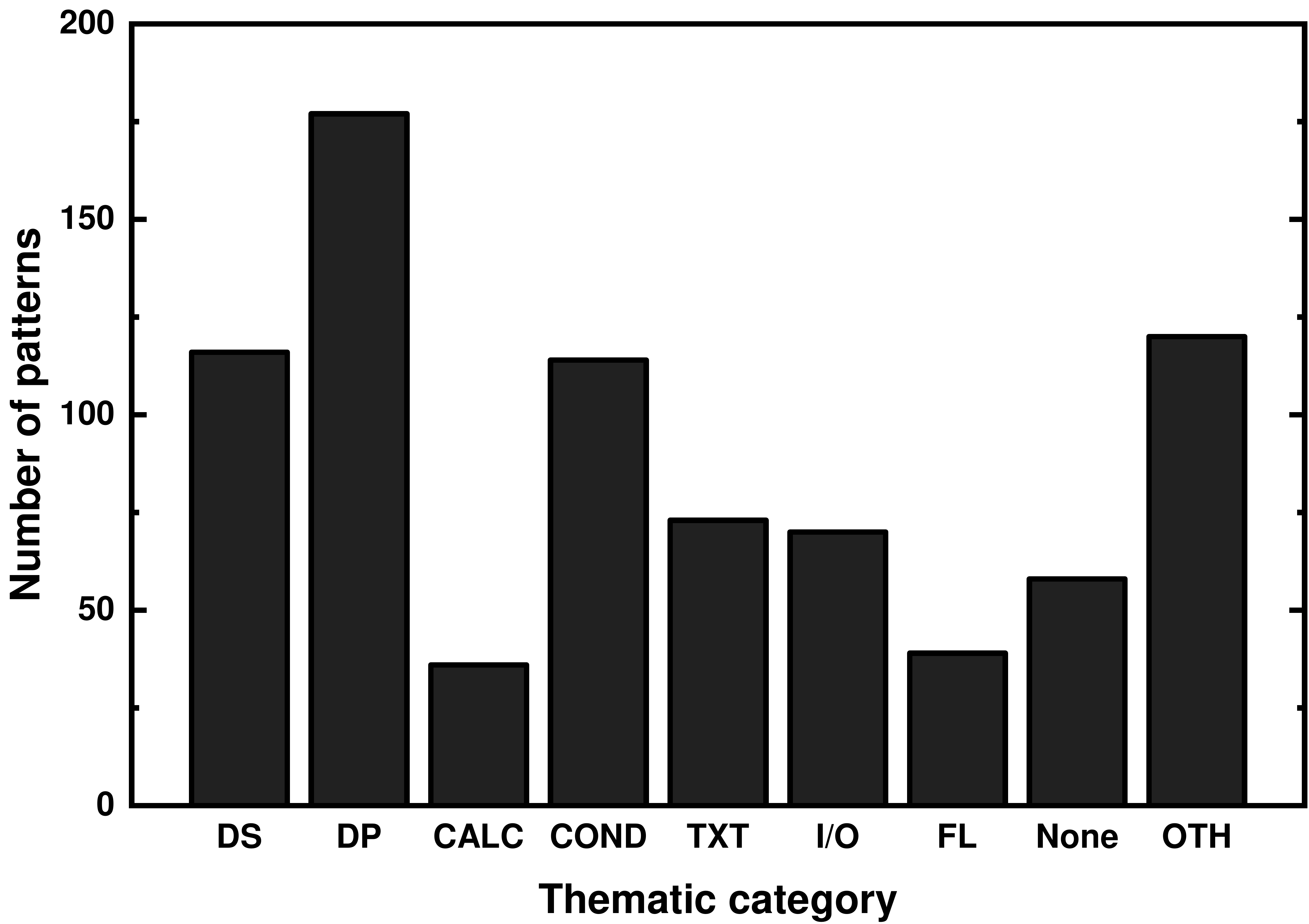}
\vspace{-0.2cm}
\caption{The distribution of patterns by their topic. \textit{DS} stands for \textit{Data Structures}, \textit{DP} -- \textit{Data Processing}, \textit{CALC} -- \textit{Calculations}, \textit{COND} -- \textit{Conditions}, \textit{TXT} -- \textit{Text}, \textit{I/O} -- \textit{Input/Output}, \textit{FL} -- \textit{Files}, \textit{OTH} -- \textit{Other}}.
\vspace{-0.4cm}
\label{fig:topical}

\end{figure}

Figure~\ref{fig:topical} shows the distribution of the patterns by their thematic categories.

\observation{The most popular thematic categories of changes are Data Processing, Data Structures, Conditions, and Other.}

\textit{Data Processing} often involves reshaping arrays, iterating over data, and working with objects. Within \textit{Data Structures}, people often change ways of working with arrays and lists, as well as matrices. As for \textit{Conditions}, there are individual interesting changes like the above-mentioned \texttt{os.path.exists()}, but the majority of them are changes in assertions within tests. One reason for the popularity of \textit{Conditions} might be that tests unite all well-designed projects from any domain, and therefore changes to the testing logic can be expected in all such repositories.

\begin{figure}[h]
\centering
\includegraphics[width=3.0in]{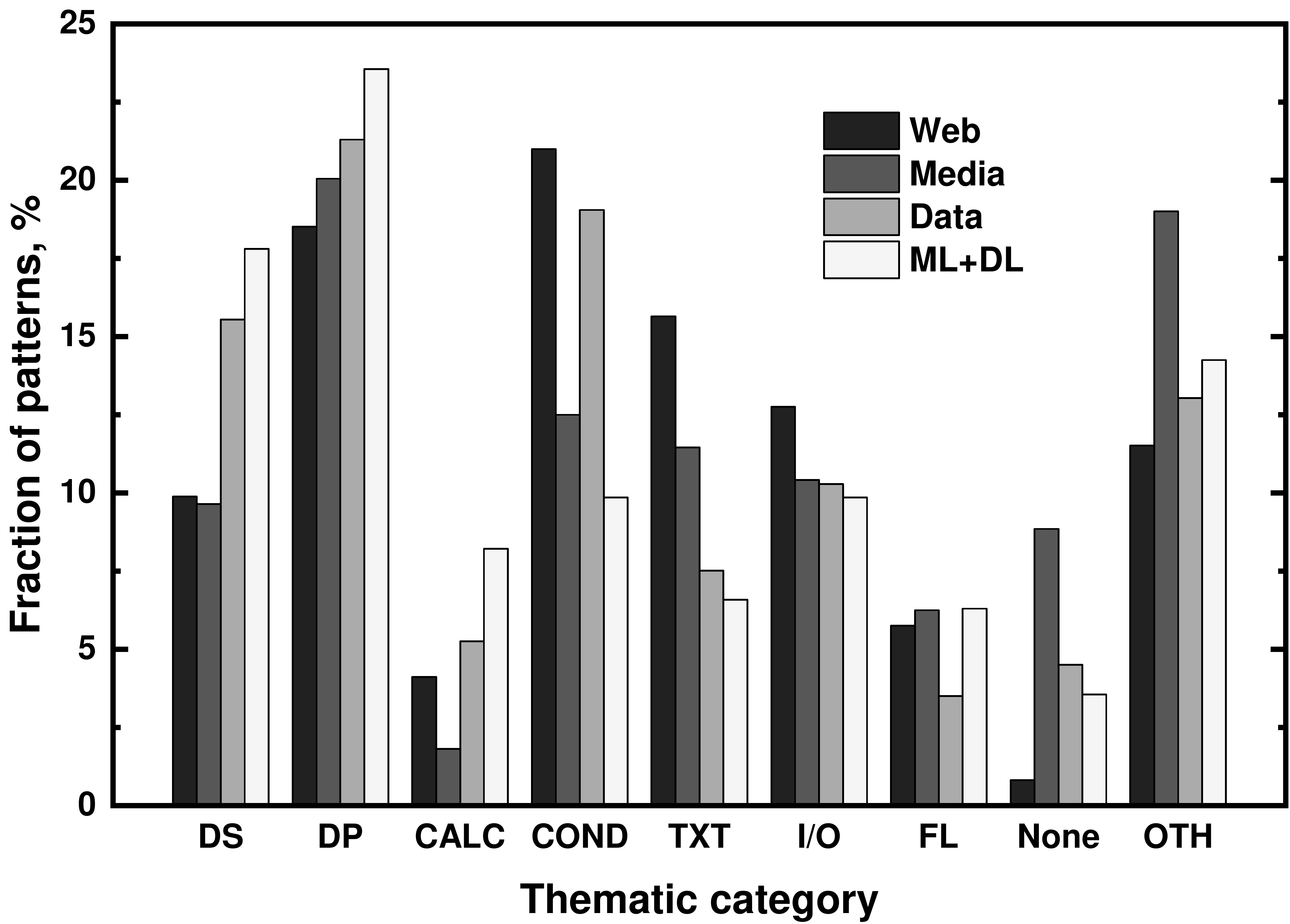}
\vspace{-0.2cm}
\caption{The distribution of patterns by their topic in different domains. \textit{DS} stands for \textit{Data Structures}, \textit{DP} -- \textit{Data Processing}, \textit{CALC} -- \textit{Calculations}, \textit{COND} -- \textit{Conditions}, \textit{TXT} -- \textit{Text}, \textit{I/O} -- \textit{Input/Output}, \textit{FL} -- \textit{Files}, \textit{OTH} -- \textit{Other}}
\vspace{-0.4cm}
\label{fig:topical_domains}
\end{figure}

Figure~\ref{fig:topical_domains} shows the distribution among the topics within different domains. It can be seen that the variation here is even more significant than for the structural categories.

Domains can be distinguished by the topical distribution within them. We can see that \textit{Data Structures} changes are more prevalent in Data and ML+DL, \textit{Calculations} come mostly from ML+DL. On the other hand, Web has the largest percentage of \textit{Text} and \textit{I/O} and Media has the largest percentage of \textit{None} and \textit{Other}. The reasons for that include a lot of specific features and a lot of pure refactorings including the graphical external libraries.

Overall, the changes in the Python code are very diverse and depend on the context of the repository and its domain. A lot of these changes have a meaningful reason behind them and are general enough to be automated or suggested to the developers in an IDE.

\subsection{GitHub Authors Survey}

To get a deeper insight into the changes that we discovered, we conducted a survey of their authors on GitHub. Each participant was asked about a change they made and also about the five changes that we picked from the most popular ones (see Section~\ref{sec:methdology_gitsurvey}). According to the responses, 85.5\% of the respondents have more than 5 years of development behind them, and 60.5\% have more than 10. 

The main results of the survey are presented in Table~\ref{table:survey}.

\begin{table}[h]
\centering
\vspace{-0.2cm}
\caption{The results of the GitHub authors survey.}
\vspace{-0.2cm}
\label{table:survey}
\begin{tabular}{C{0.22\textwidth} C{0.22\textwidth}}
\toprule
\toprule
\multicolumn{2}{c}{\textbf{Can you give your change a name?}} \\
\midrule
\textbf{Yes}  & \textbf{No} \\
\midrule
82.9\% & 17.1\%    \\
\end{tabular}
\begin{tabular}{C{0.139\textwidth} C{0.139\textwidth} C{0.139\textwidth}}
\toprule
\toprule
\multicolumn{3}{c}{\textbf{Would you like to have this change automated by a tool}} \\
\multicolumn{3}{c}{\textbf{(no matter how difficult that automation might be)?}} \\
\midrule
\textbf{Yes}  & \textbf{No} & \textbf{Already automated} \\
\midrule
57.9\% & 36.8\% & 5.3\%   \\
\bottomrule
\end{tabular}
\end{table}

The answers to the first question show that the vast majority of the respondents can give their change a name. The given names include \textit{API change}, \textit{Updating a deprecated usage}, \textit{Switch from Pillow to openCV}. As for the respondents who were unable to provide a name for the change, they listed several reasons for it. Several authors reported that their change was too specific, others that their change was a part of a larger one and therefore highly depended on the surrounding logic. Overall, most authors gave their change a concise and meaningful name indicating that this change did have a specific purpose behind it.

The answers to the second question show that more than a half of our respondents would like to have their previous change automated. Due to the small number of emails and a large number of patterns, we only had 4 cases where two people evaluated changes in the same patterns. In 3 of such cases the respondents agreed between each other, and in the fourth case, one author said that they would like the deprecated alias of a \texttt{unittest} method \texttt{assertNotEquals} to be automatically changed to a standard name \texttt{assertNotEqual}, while the other one said that search and replace works for them.

Other authors also elaborated on the reasons why they would not like their changes automated. Some changes are too specific or too dependant on the context around them, some changes implement a new feature, and some developers simply do not trust the automatic refactoring systems and want to have the full manual control over their work. Still, the majority of the authors expressed the desire to have their change automated.

\begin{figure}[h]
\centering
\vspace{-0.2cm}
\includegraphics[width=3in]{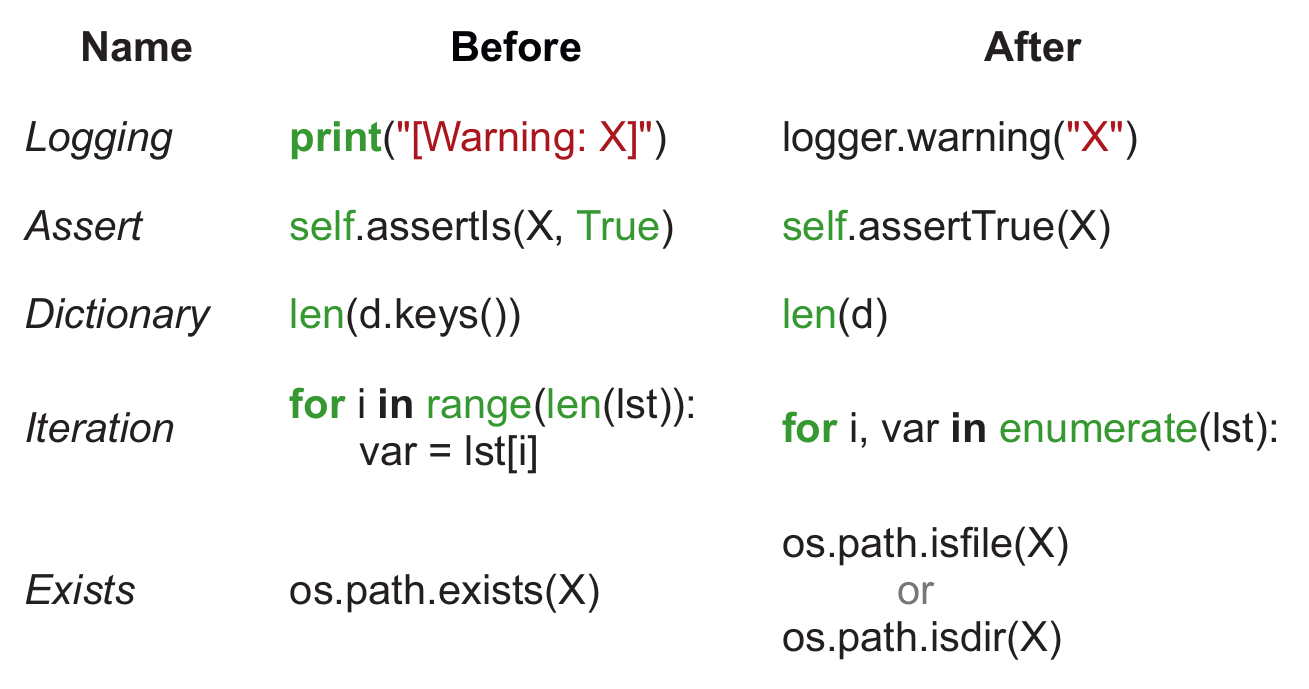}
\vspace{-0.2cm}
\caption{Five of the most popular change patterns given to the participants of the GitHub survey to rank.}
\vspace{-0.2cm}
\label{fig:survey_patterns}
\end{figure}

Besides their own change, the participants were asked to rank 5 selected changes (see Section~\ref{sec:categorization}) from the standpoint of the usefulness of their automation. The chosen patterns are presented in Figure~\ref{fig:survey_patterns}. 

\begin{figure}[h]
\centering
\includegraphics[width=3.0in]{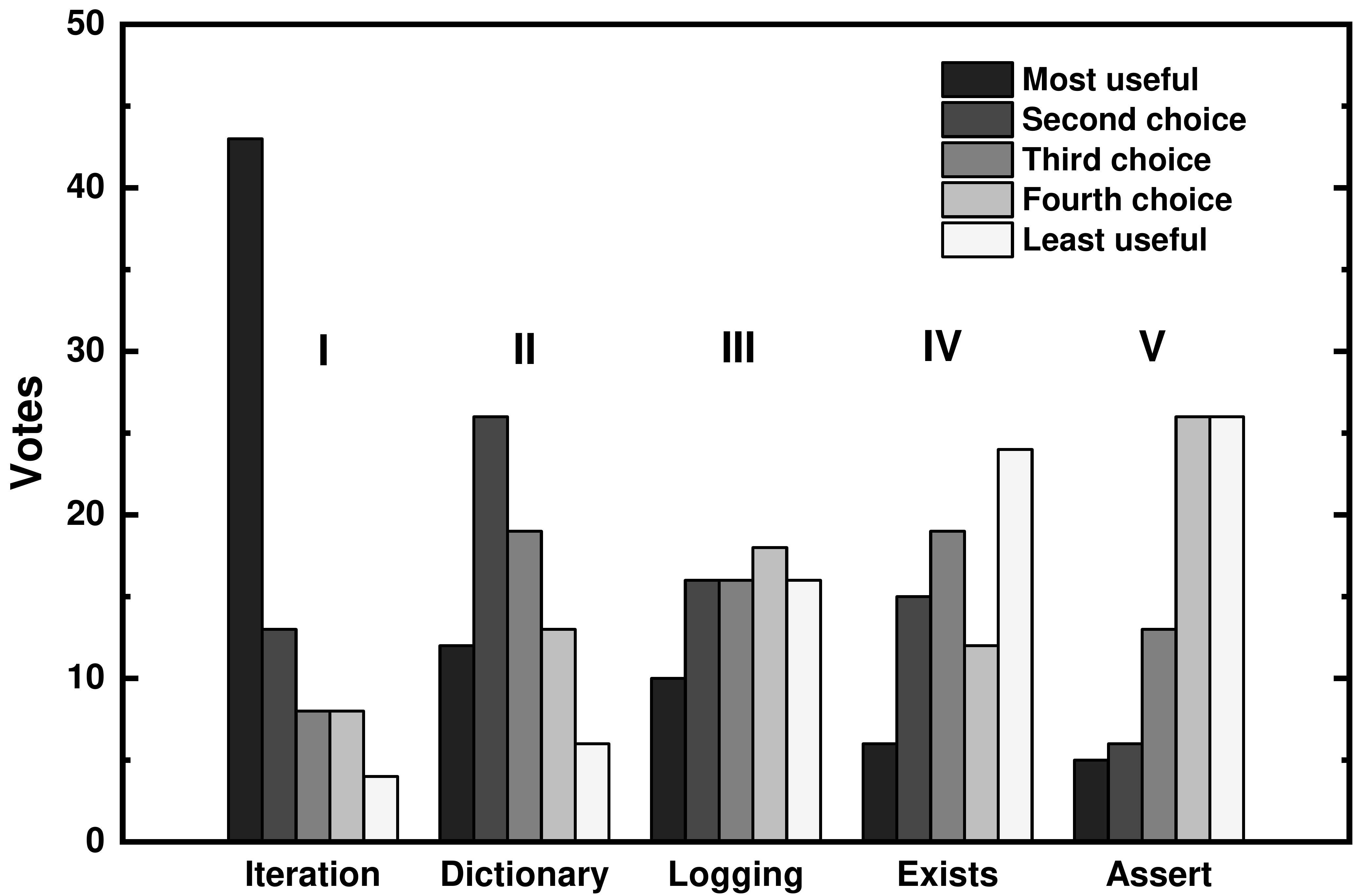}
\caption{The votes of the participants of the GitHub survey on the usefulness of automating the suggested changes.}
\label{fig:five_results}
\end{figure}

The results of the vote are presented in Figure~\ref{fig:five_results}. Counting the \textit{Most useful} as 5, \textit{Second choice} as 4, and so on, we calculated the average score of each change pattern and compiled their final ranking, shown in Roman numerals above each change.

Firstly, we can notice a very large breakaway of the most popular answer. \textit{Iteration}, while being probably the most conceptually complex of all the ranked patterns, has a striking lead, with 56.6\% of respondents considering its automation to be the most useful. This once again shows the usefulness of \toolname for discovering useful popular changes as it can discover changes that contain several parts separated by other lines of code and are discovered specifically on the fine-grained program dependence graph.

Another interesting feature of this ranking is that two patterns with the most positive feedback, \textit{Iteration} and \textit{Dictionary}, are also the only ones that deal with Built-in Functions, which may indicate developers' preference towards such changes. On the other hand, the least popular changes both occur in libraries.

\subsection{IDE Team Interview}

The final stage of our study is interviewing the members of the IDE development team to understand whether automating the discovered changes is useful and realistic.

Firstly, to interview the PyCharm team, we needed to select more patterns that have the potential to be automated from those approved by their authors in the survey. In total, we received 44 positive votes for automation in the GitHub authors survey that cover 43 patterns (two votes refer to two samples in the same pattern). We applied open coding as described in Section~\ref{sec:methdology_ide} and obtained the following groups of patterns:

\begin{itemize}
    \item \textbf{Renames} --- patterns that a lot of people ask to have automated even though they can be automated using existing IDE features.
    \item \textbf{Deprecations} --- patterns that update deprecated APIs, and might not be of interest at present.
    \item \textbf{New Features} --- patterns that implement entirely new features and are therefore impossible to automate.
    \item \textbf{Specific Code} --- patterns that are too specific for general automations.
    \item \textbf{Potential Automations} --- patterns that do not fall into the described groups and can therefore be considered as candidates for possible automation.
\end{itemize}

\begin{table}[h]
\centering
\vspace{-0.2cm}
\caption{The categories of the patterns that received a positive feedback from their GitHub authors in the survey. Deprec. stands for Deprecations.}
\vspace{-0.2cm}
\label{table:survey_good}
\begin{tabular}{c c c c c}
\toprule
\multirow{2}{*}{\textbf{Renames}} & \multirow{2}{*}{\textbf{Deprec.}} & \textbf{New} & \textbf{Specific} & \textbf{Potential} \\
                  &                   & \textbf{Features} & \textbf{Code} & \textbf{Automations} \\
\midrule
10 & 5 & 1 & 19 & \textbf{8}\\
\bottomrule
\vspace{-0.4cm}
\end{tabular}
\end{table}
The distribution of these categories within our 43 patterns is presented in Table~\ref{table:survey_good}. It can be seen that a large portion of the GitHub authors wants to automate various types of renamings of functions (that vary in difficulty but can be generally covered with existing IDE features) and features that are very specific like changing a specific string to a variable or moving a logarithm from natural to base 2. In total, 8 patterns were general enough to consider turning them into automations, of which we selected 5. Together with the 10 patterns chosen after the categorization (see Section~\ref{sec:categorization}), this comprises 15 patterns that we showed to IDE team members. The names and full descriptions of the chosen patterns are available.\footnote{15 chosen patterns: \url{https://zenodo.org/record/4004174}}

\begin{figure}[h]
\vspace{-0.2cm}
\centering
\includegraphics[width=3in]{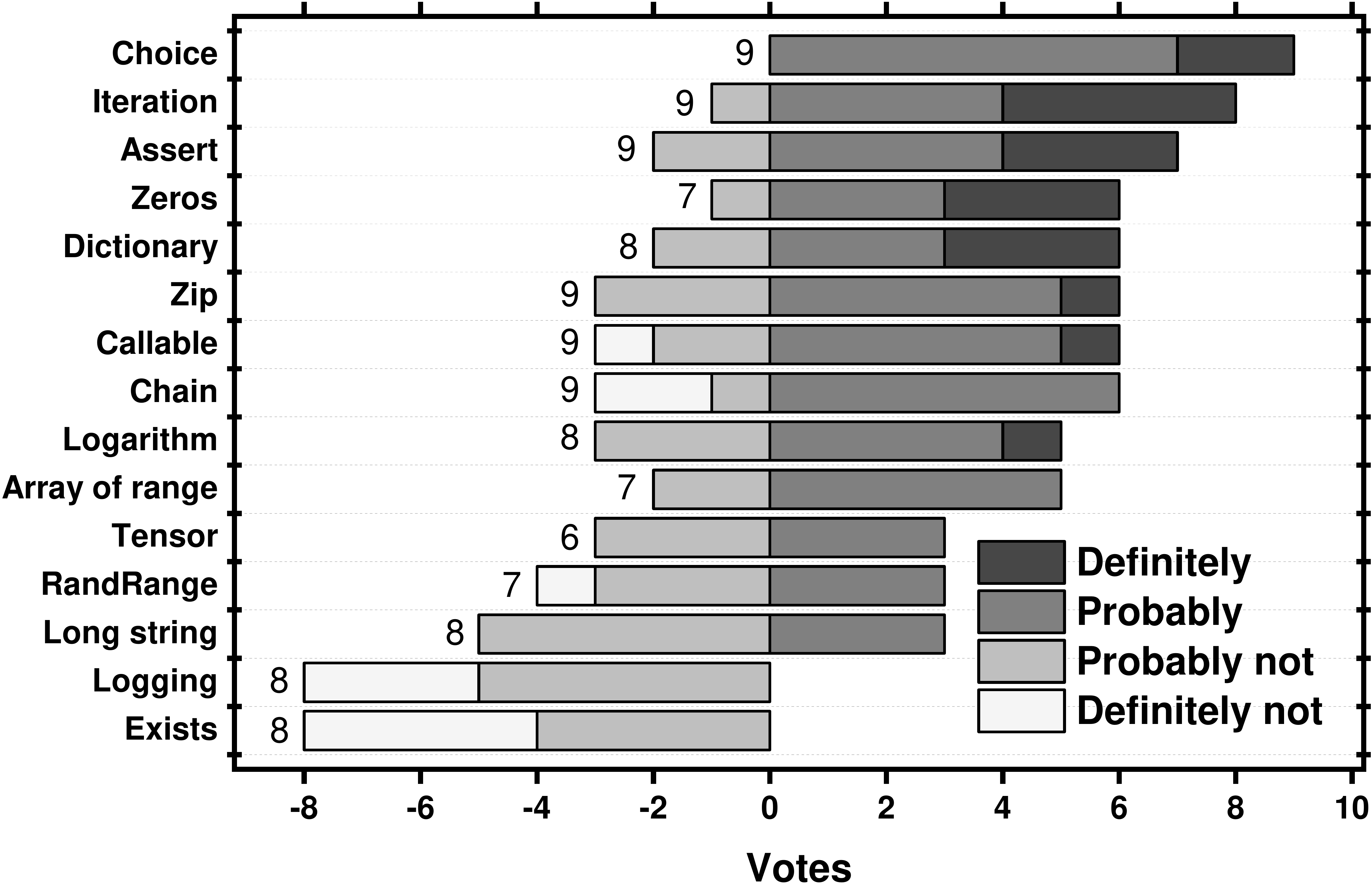}
\vspace{-0.2cm}
\caption{The votes of the PyCharm team on the usefulness of automating the suggested changes.}
\vspace{-0.2cm}
\label{fig:pycharm}
\end{figure}

The first and the main question of the interview was \textit{Do you think that this change should be suggested to the user via an inspection or an intention in PyCharm?} The distribution of the opinions of the team is presented in Figure~\ref{fig:pycharm}.

In total, we interviewed 9 candidates, but because of the possible \textit{Don't know} answer, different patterns have a different number of votes, indicated by the number to the left of each bar.

It can be seen that a lot of changes received positive feedback from the PyCharm team. Comparing the opinions with the GitHub authors, there is one significant difference. \textit{Iteration} and \textit{Dictionary} are once again among the most popular patterns. The IDE team member S7 described \textit{Iteration} like this: \textit{``An inspection like this can help beginners to learn some language features (e.g. enumerate) and make code more idiomatic.''} On the other end of the spectrum, \textit{Logging} and \textit{Exists} are at the very bottom, without a single positive vote. However, \textit{Assert}, which was the last in the GitHub authors survey, is among the best ones here. S2 simply described \textit{Assert} as \textit{``Makes code much nicer''}.

To understand the difference, firstly, we can use the answers to the second question that concerns the difficulty of automating the patterns. While in general we found no specific correlation between the answers to these questions, both \textit{Logging} and \textit{Exists} are among the patterns that the PyCharm team considered to be the hardest to automate. Meanwhile, \textit{Assert} is directly opposite, among the patterns where the majority of IDE team members chose the \textit{Simple} answer. Secondly, we can look at the commentaries that the team gave in the third question. They show that \textit{Logging} and \textit{Exists} require too much context or may produce too many false positives. The respondent S2 said about \textit{Exists}, \textit{``We don't know what the developer wants here''}. They also commented on \textit{Logging}, \textit{``We don't know how logging is organized in a specific program''}.

\observation{The PyCharm development team considers that popular changes may be automated if these changes are simple, precise, and require little context.}

Another problem that concerned the interviewees was being able to accurately determine the users who would need specific suggestions. For example, fixing deprecated APIs or changes related to scientific calculations can be of interest to some users and only bother others. Researching new ways of targeting suggestions seems very relevant and important in this regard.

Among the other changes that received positive votes is \textit{Zeros}, which consists of moving from \texttt{numpy.array([0, 0, 0...])} to \texttt{numpy.zeros(X)}. Interestingly though, the change that came out first and did not receive a single negative vote actually comes from the answers to the GitHub authors survey. \textit{Choice} relates to a module \texttt{random} from the Standard Library and moves from:

\hspace{0.6cm}\texttt{i = random.randrange(0, len(X))}

\hspace{0.6cm}\texttt{var = x[i]} 

towards

\hspace{0.6cm}\texttt{var = random.choice(X)}

to directly pick a random item in the list. This makes the code more concise and readable.

Overall, \toolname allowed us to discover popular changes with a high potential of being automated for the convenience of the developer, and several changes received positive feedback from GitHub authors and the members of the PyCharm development team. In the future this tool can be used on a larger scale to discover even more useful patterns.

%% file: sections/06-discussion.tex
\section{Implications}
\label{sec:discussion}

\textbf{Implications for tool builders}. We identified changes that received positive feedback both from the GitHub authors and from the IDE development team from the standpoint of automation. \textit{Iteration} and \textit{Dictionary}, for example, can be added to IDE suggestions. Secondly, we provided insights into the mindsets of developers and IDE creators that may also be useful to understand what people want. And finally, our tool can serve as the first step towards the future systems of continuous monitoring that mines the projects on a regular basis and incorporates the newly detected patterns into the IDE.

\textbf{Implications for researchers}. In our work, we used fgPDG for the first time for Python and also investigated the usefulness of discovered patterns for automation. Both of these aspects of the study can be continued on a larger scale, to discover more insights into the nature of code evolution and also to gather more patterns that show potential for automation. Future research may also focus on specific types of changes (for example, bug-fixing or refactorings) or specific domains. Specifically, it might be of interest to study ML projects more closely, because not all researchers and data scientists from ML are expert programmers, and potential suggestions could be beneficial for them. Also, interviewing the IDE development team demonstrated that there exists a real need of researching ways of making sure that a specific suggestion or automation is suitable for a specific user.

\textbf{Implications for developers}. During the GitHub authors survey, it appeared that a lot of automation requests were for already automated features available in popular IDEs. This lack of awareness of existing features is a known problem~\cite{murphy2012improving}, meaning that developers should pay closer attention to the tooltips in the IDE and that IDEs should reach out to their users more.

%% file: sections/07-threats.tex
\vspace{-0.2cm}
\section{Threats to Validity}
\label{sec:threats}

We have taken care to ensure that our results are unbiased, and in this section, we discuss threats to validity for our study and the ways we mitigated them.

One of the threats to validity is related to the generalization of our findings. Even though we collected a moderately large dataset of projects and selected them from different domains, our observations still relate only to these specific repositories. Also we considered only GitHub as a platform for open-source software projects. However, we believe the changes identified through our study are useful regardless of the platform they came from.

The primary threats to the internal validity of this study are possible faults in the implementation of \toolname. We control this threat by extensively testing our implementations and verifying their results against a smaller dataset for which we can manually determine the correct results. Another threat is the limited capability of \toolname as it does not support all the language constructs within Python during parsing and mapping. For example, \textit{finally} branches are not supported, as well as generators. Moreover, in this work, we do not distinguish between Python versions 2 and 3. The tool was written to support Python 3.8, however, it can parse Python 2 functions that have grammar constructs compatible with Python 3.

Although these threats are important to note, we believe they do not invalidate the main findings of our research. \toolname can be useful for a number of different tasks, and the patterns that we studied can be employed for improving existing IDEs.
\vspace{-0.2cm}

%% file: sections/08-conclusion.tex
\section{Conclusion}
\label{sec:conclusions}

In this paper, we conducted a study of code change patterns in Python projects hosted on GitHub and researched the possibility of their automation.

We developed a tool called \toolname that searches for code change patterns in Python. By running the tool on a dataset of 120 GitHub projects, we discovered 7,481 patterns. Of these, 803 were cross-project. 

Our survey of GitHub authors regarding their opinion about the possible automation of the discovered patterns revealed that majority (57.9\%) of the participants would like to have their change automated. When asked about the perceived usefulness of automating five popular changes, respondents preferred the changes to built-in functions over the changes to functions from popular libraries. The most popular change, \textit{Iteration}, highlights the ability of the tool to discover changes that occur over several lines of code.

In the end, to understand the practical possibility of automating the discovered changes, we interviewed nine members of the development team of PyCharm. Several of the discovered change patterns received favorable votes from the PyCharm team and can be considered for automation for any IDE out there. \textit{Choice}, \textit{Iteration}, \textit{Assert}, \textit{Zeros}, and \textit{Dictionary} are among the top ones.

Finally, to answer the question that the title of our paper poses. These changes from the trenches --- should we automate them? According to our study, the answer seems to be: yes, but not all of them. A significant percentage of developers in the survey said that they did not want their changes to be automated, whereas a lot of them said they would. Additionally, at least one pattern in our study received almost opposite opinions from the GitHub authors and the IDE team members. The IDE team also shared several specific prerequisites for automating such changes. That said, several patterns received largely positive votes by both parties, enabling us to conclude that mining frequent changes from GitHub is useful for finding possible automations, and more future work should be conducted in this direction. 

\section*{Acknowledgements}
This research was supported in part through computational resources of HPC facilities at NRU HSE.

%% file: main.bbl

\begin{thebibliography}{40}

%%% the Web address.

\ifx \showCODEN    \undefined \def \showCODEN     #1{\unskip}     \fi
\ifx \showDOI      \undefined \def \showDOI       #1{#1}\fi
\ifx \showISBNx    \undefined \def \showISBNx     #1{\unskip}     \fi
\ifx \showISBNxiii \undefined \def \showISBNxiii  #1{\unskip}     \fi
\ifx \showISSN     \undefined \def \showISSN      #1{\unskip}     \fi
\ifx \showLCCN     \undefined \def \showLCCN      #1{\unskip}     \fi
\ifx \shownote     \undefined \def \shownote      #1{#1}          \fi
\ifx \showarticletitle \undefined \def \showarticletitle #1{#1}   \fi
\ifx \showURL      \undefined \def \showURL       {\relax}        \fi
% invisible to TeX
\providecommand\bibfield[2]{#2}
\providecommand\bibinfo[2]{#2}
\providecommand\natexlab[1]{#1}
\providecommand\showeprint[2][]{arXiv:#2}

\bibitem[\protect\citeauthoryear{Chakraborty, Baowaly, Arefin, and
  Bahar}{Chakraborty et~al\mbox{.}}{2012}]%
        {chakraborty2012role}
\bibfield{author}{\bibinfo{person}{Abhijit Chakraborty},
  \bibinfo{person}{Mrinal~Kanti Baowaly}, \bibinfo{person}{Ashraful Arefin},
  {and} \bibinfo{person}{Ali~Newaz Bahar}.} \bibinfo{year}{2012}\natexlab{}.
\newblock \showarticletitle{The role of requirement engineering in software
  development life cycle}.
\newblock \bibinfo{journal}{\emph{Journal of emerging trends in computing and
  information sciences}} \bibinfo{volume}{3}, \bibinfo{number}{5}
  (\bibinfo{year}{2012}), \bibinfo{pages}{723--729}.
\newblock


\bibitem[\protect\citeauthoryear{Chen, Ma, Lin, Chen, Li, and Xu}{Chen
  et~al\mbox{.}}{2018}]%
        {chen2018study}
\bibfield{author}{\bibinfo{person}{Zhifei Chen}, \bibinfo{person}{Wanwangying
  Ma}, \bibinfo{person}{Wei Lin}, \bibinfo{person}{Lin Chen},
  \bibinfo{person}{Yanhui Li}, {and} \bibinfo{person}{Baowen Xu}.}
  \bibinfo{year}{2018}\natexlab{}.
\newblock \showarticletitle{A study on the changes of dynamic feature code when
  fixing bugs: towards the benefits and costs of Python dynamic features}.
\newblock \bibinfo{journal}{\emph{Science China Information Sciences}}
  \bibinfo{volume}{61}, \bibinfo{number}{1} (\bibinfo{year}{2018}),
  \bibinfo{pages}{012107}.
\newblock


\bibitem[\protect\citeauthoryear{Cotroneo, De~Simone, Iannillo, Natella,
  Rosiello, and Bidokhti}{Cotroneo et~al\mbox{.}}{2019}]%
        {cotroneo2019analyzing}
\bibfield{author}{\bibinfo{person}{Domenico Cotroneo}, \bibinfo{person}{Luigi
  De~Simone}, \bibinfo{person}{Antonio~Ken Iannillo}, \bibinfo{person}{Roberto
  Natella}, \bibinfo{person}{Stefano Rosiello}, {and}
  \bibinfo{person}{Nematollah Bidokhti}.} \bibinfo{year}{2019}\natexlab{}.
\newblock \showarticletitle{Analyzing the context of bug-fixing changes in the
  openstack cloud computing platform}. In \bibinfo{booktitle}{\emph{2019 IEEE
  30th International Symposium on Software Reliability Engineering (ISSRE)}}.
  \bibinfo{pages}{334--345}.
\newblock


\bibitem[\protect\citeauthoryear{Cupy}{Cupy}{2021}]%
        {cupy_commit}
\bibfield{author}{\bibinfo{person}{Cupy}.} \bibinfo{year}{accessed:
  01.08.2021}\natexlab{}.
\newblock \bibinfo{title}{A commit in Cupy}.
\newblock
\newblock
\urldef\tempurl%
\url{https://github.com/cupy/cupy/commit/3d1cdc43e7df411ee51bfaa516356c668e4c2f6d}
\showURL{%
\tempurl}


\bibitem[\protect\citeauthoryear{Dagenais and Robillard}{Dagenais and
  Robillard}{2011}]%
        {dagenais2011recommending}
\bibfield{author}{\bibinfo{person}{Barth{\'e}l{\'e}my Dagenais} {and}
  \bibinfo{person}{Martin~P Robillard}.} \bibinfo{year}{2011}\natexlab{}.
\newblock \showarticletitle{Recommending adaptive changes for framework
  evolution}.
\newblock \bibinfo{journal}{\emph{ACM Transactions on Software Engineering and
  Methodology (TOSEM)}} \bibinfo{volume}{20}, \bibinfo{number}{4}
  (\bibinfo{year}{2011}), \bibinfo{pages}{1--35}.
\newblock


\bibitem[\protect\citeauthoryear{Falleri, Morandat, Blanc, Martinez, and
  Monperrus}{Falleri et~al\mbox{.}}{2014}]%
        {DBLP:conf/kbse/FalleriMBMM14}
\bibfield{author}{\bibinfo{person}{Jean{-}R{\'{e}}my Falleri},
  \bibinfo{person}{Flor{\'{e}}al Morandat}, \bibinfo{person}{Xavier Blanc},
  \bibinfo{person}{Matias Martinez}, {and} \bibinfo{person}{Martin Monperrus}.}
  \bibinfo{year}{2014}\natexlab{}.
\newblock \showarticletitle{Fine-grained and accurate source code
  differencing}. In \bibinfo{booktitle}{\emph{{ACM/IEEE} International
  Conference on Automated Software Engineering, {ASE} '14, Vasteras, Sweden -
  September 15 - 19, 2014}}. \bibinfo{pages}{313--324}.
\newblock


\bibitem[\protect\citeauthoryear{Garousi and Felderer}{Garousi and
  Felderer}{2017}]%
        {garousi2017worlds}
\bibfield{author}{\bibinfo{person}{Vahid Garousi} {and}
  \bibinfo{person}{Michael Felderer}.} \bibinfo{year}{2017}\natexlab{}.
\newblock \showarticletitle{Worlds apart: industrial and academic focus areas
  in software testing}.
\newblock \bibinfo{journal}{\emph{IEEE Software}} \bibinfo{volume}{34},
  \bibinfo{number}{5} (\bibinfo{year}{2017}), \bibinfo{pages}{38--45}.
\newblock


\bibitem[\protect\citeauthoryear{German}{German}{2006}]%
        {german2006empirical}
\bibfield{author}{\bibinfo{person}{Daniel~M German}.}
  \bibinfo{year}{2006}\natexlab{}.
\newblock \showarticletitle{An empirical study of fine-grained software
  modifications}.
\newblock \bibinfo{journal}{\emph{Empirical Software Engineering}}
  \bibinfo{volume}{11}, \bibinfo{number}{3} (\bibinfo{year}{2006}),
  \bibinfo{pages}{369--393}.
\newblock


\bibitem[\protect\citeauthoryear{Hassan}{Hassan}{2009}]%
        {hassan2009predicting}
\bibfield{author}{\bibinfo{person}{Ahmed~E Hassan}.}
  \bibinfo{year}{2009}\natexlab{}.
\newblock \showarticletitle{Predicting faults using the complexity of code
  changes}. In \bibinfo{booktitle}{\emph{2009 IEEE 31st international
  conference on software engineering}}. \bibinfo{pages}{78--88}.
\newblock


\bibitem[\protect\citeauthoryear{Kalliamvakou, Gousios, Blincoe, Singer,
  German, and Damian}{Kalliamvakou et~al\mbox{.}}{2014}]%
        {kalliamvakou2014promises}
\bibfield{author}{\bibinfo{person}{Eirini Kalliamvakou},
  \bibinfo{person}{Georgios Gousios}, \bibinfo{person}{Kelly Blincoe},
  \bibinfo{person}{Leif Singer}, \bibinfo{person}{Daniel~M German}, {and}
  \bibinfo{person}{Daniela Damian}.} \bibinfo{year}{2014}\natexlab{}.
\newblock \showarticletitle{The promises and perils of mining github}. In
  \bibinfo{booktitle}{\emph{Proceedings of the 11th working conference on
  mining software repositories}}. \bibinfo{pages}{92--101}.
\newblock


\bibitem[\protect\citeauthoryear{Kawrykow and Robillard}{Kawrykow and
  Robillard}{2011}]%
        {kawrykow2011non}
\bibfield{author}{\bibinfo{person}{David Kawrykow} {and}
  \bibinfo{person}{Martin~P Robillard}.} \bibinfo{year}{2011}\natexlab{}.
\newblock \showarticletitle{Non-essential changes in version histories}. In
  \bibinfo{booktitle}{\emph{2011 33rd International Conference on Software
  Engineering (ICSE)}}. \bibinfo{pages}{351--360}.
\newblock


\bibitem[\protect\citeauthoryear{Kim, Whitehead, and Zhang}{Kim
  et~al\mbox{.}}{2008}]%
        {kim2008classifying}
\bibfield{author}{\bibinfo{person}{Sunghun Kim}, \bibinfo{person}{E~James
  Whitehead}, {and} \bibinfo{person}{Yi Zhang}.}
  \bibinfo{year}{2008}\natexlab{}.
\newblock \showarticletitle{Classifying software changes: Clean or buggy?}
\newblock \bibinfo{journal}{\emph{IEEE Transactions on Software Engineering}}
  \bibinfo{volume}{34}, \bibinfo{number}{2} (\bibinfo{year}{2008}),
  \bibinfo{pages}{181--196}.
\newblock


\bibitem[\protect\citeauthoryear{Kim, Zimmermann, Pan, James~Jr,
  et~al\mbox{.}}{Kim et~al\mbox{.}}{2006}]%
        {kim2006automatic}
\bibfield{author}{\bibinfo{person}{Sunghun Kim}, \bibinfo{person}{Thomas
  Zimmermann}, \bibinfo{person}{Kai Pan}, \bibinfo{person}{E James~Jr},
  {et~al\mbox{.}}} \bibinfo{year}{2006}\natexlab{}.
\newblock \showarticletitle{Automatic identification of bug-introducing
  changes}. In \bibinfo{booktitle}{\emph{21st IEEE/ACM international conference
  on automated software engineering (ASE'06)}}. \bibinfo{pages}{81--90}.
\newblock


\bibitem[\protect\citeauthoryear{Koyuncu, Liu, Bissyand{\'e}, Kim, Klein,
  Monperrus, and Le~Traon}{Koyuncu et~al\mbox{.}}{2020}]%
        {koyuncu2020fixminer}
\bibfield{author}{\bibinfo{person}{Anil Koyuncu}, \bibinfo{person}{Kui Liu},
  \bibinfo{person}{Tegawend{\'e}~F Bissyand{\'e}}, \bibinfo{person}{Dongsun
  Kim}, \bibinfo{person}{Jacques Klein}, \bibinfo{person}{Martin Monperrus},
  {and} \bibinfo{person}{Yves Le~Traon}.} \bibinfo{year}{2020}\natexlab{}.
\newblock \showarticletitle{Fixminer: Mining relevant fix patterns for
  automated program repair}.
\newblock \bibinfo{journal}{\emph{Empirical Software Engineering}}
  (\bibinfo{year}{2020}), \bibinfo{pages}{1--45}.
\newblock


\bibitem[\protect\citeauthoryear{Lenarduzzi, Sillitti, and Taibi}{Lenarduzzi
  et~al\mbox{.}}{2017}]%
        {lenarduzzi2017analyzing}
\bibfield{author}{\bibinfo{person}{Valentina Lenarduzzi},
  \bibinfo{person}{Alberto Sillitti}, {and} \bibinfo{person}{Davide Taibi}.}
  \bibinfo{year}{2017}\natexlab{}.
\newblock \showarticletitle{Analyzing forty years of software maintenance
  models}. In \bibinfo{booktitle}{\emph{2017 IEEE/ACM 39th International
  Conference on Software Engineering Companion (ICSE-C)}}.
  \bibinfo{pages}{146--148}.
\newblock


\bibitem[\protect\citeauthoryear{Lin, Chen, Ma, Chen, Xu, and Xu}{Lin
  et~al\mbox{.}}{2016}]%
        {lin2016empirical}
\bibfield{author}{\bibinfo{person}{Wei Lin}, \bibinfo{person}{Zhifei Chen},
  \bibinfo{person}{Wanwangying Ma}, \bibinfo{person}{Lin Chen},
  \bibinfo{person}{Lei Xu}, {and} \bibinfo{person}{Baowen Xu}.}
  \bibinfo{year}{2016}\natexlab{}.
\newblock \showarticletitle{An empirical study on the characteristics of Python
  fine-grained source code change types}. In \bibinfo{booktitle}{\emph{2016
  IEEE international conference on software maintenance and evolution
  (ICSME)}}. \bibinfo{pages}{188--199}.
\newblock


\bibitem[\protect\citeauthoryear{Martinez, Duchien, and Monperrus}{Martinez
  et~al\mbox{.}}{2014}]%
        {martinez2014accurate}
\bibfield{author}{\bibinfo{person}{Matias Martinez}, \bibinfo{person}{Laurence
  Duchien}, {and} \bibinfo{person}{Martin Monperrus}.}
  \bibinfo{year}{2014}\natexlab{}.
\newblock \showarticletitle{Accurate Extraction of Bug Fix Pattern Occurrences
  using Abstract Syntax Tree Analysis}.
\newblock  (\bibinfo{year}{2014}).
\newblock


\bibitem[\protect\citeauthoryear{Medeiros, Ribeiro, Gheyi, Apel, K{\"{a}}stner,
  Ferreira, Carvalho, and Fonseca}{Medeiros et~al\mbox{.}}{2018}]%
        {FlavioMedeiros_2018}
\bibfield{author}{\bibinfo{person}{Fl{\'{a}}vio Medeiros},
  \bibinfo{person}{M{\'{a}}rcio Ribeiro}, \bibinfo{person}{Rohit Gheyi},
  \bibinfo{person}{Sven Apel}, \bibinfo{person}{Christian K{\"{a}}stner},
  \bibinfo{person}{Bruno Ferreira}, \bibinfo{person}{Luiz Carvalho}, {and}
  \bibinfo{person}{Baldoino Fonseca}.} \bibinfo{year}{2018}\natexlab{}.
\newblock \showarticletitle{Discipline Matters: Refactoring of Preprocessor
  Directives in the {\#}ifdef Hell}.
\newblock \bibinfo{journal}{\emph{{IEEE} Transactions on Software Engineering}}
  \bibinfo{volume}{44}, \bibinfo{number}{5} (\bibinfo{year}{2018}),
  \bibinfo{pages}{453--469}.
\newblock


\bibitem[\protect\citeauthoryear{Meqdadi and Aljawarneh}{Meqdadi and
  Aljawarneh}{2020}]%
        {meqdadi2020study}
\bibfield{author}{\bibinfo{person}{Omar Meqdadi} {and} \bibinfo{person}{Shadi
  Aljawarneh}.} \bibinfo{year}{2020}\natexlab{}.
\newblock \showarticletitle{A study of code change patterns for adaptive
  maintenance with AST analysis}.
\newblock \bibinfo{journal}{\emph{International Journal of Electrical and
  Computer Engineering}} \bibinfo{volume}{10}, \bibinfo{number}{3}
  (\bibinfo{year}{2020}), \bibinfo{pages}{2719}.
\newblock


\bibitem[\protect\citeauthoryear{Mockus and Weiss}{Mockus and Weiss}{2000}]%
        {mockus2000predicting}
\bibfield{author}{\bibinfo{person}{Audris Mockus} {and}
  \bibinfo{person}{David~M Weiss}.} \bibinfo{year}{2000}\natexlab{}.
\newblock \showarticletitle{Predicting risk of software changes}.
\newblock \bibinfo{journal}{\emph{Bell Labs Technical Journal}}
  \bibinfo{volume}{5}, \bibinfo{number}{2} (\bibinfo{year}{2000}),
  \bibinfo{pages}{169--180}.
\newblock


\bibitem[\protect\citeauthoryear{Multiflow}{Multiflow}{2021}]%
        {scikit_commit}
\bibfield{author}{\bibinfo{person}{SciKit Multiflow}.} \bibinfo{year}{accessed:
  01.08.2021}\natexlab{}.
\newblock \bibinfo{title}{A commit in SciKit Multiflow}.
\newblock
\newblock
\urldef\tempurl%
\url{https://github.com/scikit-multiflow/scikit-multiflow/commit/73b13f228b398e334e4c65ff286f5ad02932ce7f}
\showURL{%
\tempurl}


\bibitem[\protect\citeauthoryear{Murphy-Hill, Jiresal, and Murphy}{Murphy-Hill
  et~al\mbox{.}}{2012}]%
        {murphy2012improving}
\bibfield{author}{\bibinfo{person}{Emerson Murphy-Hill}, \bibinfo{person}{Rahul
  Jiresal}, {and} \bibinfo{person}{Gail~C Murphy}.}
  \bibinfo{year}{2012}\natexlab{}.
\newblock \showarticletitle{Improving software developers' fluency by
  recommending development environment commands}. In
  \bibinfo{booktitle}{\emph{Proceedings of the ACM SIGSOFT 20th International
  Symposium on the Foundations of Software Engineering}}.
  \bibinfo{pages}{1--11}.
\newblock


\bibitem[\protect\citeauthoryear{Nagpal and Gabrani}{Nagpal and
  Gabrani}{2019}]%
        {nagpal2019python}
\bibfield{author}{\bibinfo{person}{Abhinav Nagpal} {and}
  \bibinfo{person}{Goldie Gabrani}.} \bibinfo{year}{2019}\natexlab{}.
\newblock \showarticletitle{Python for data analytics, scientific and technical
  applications}. In \bibinfo{booktitle}{\emph{2019 Amity international
  conference on artificial intelligence (AICAI)}}. \bibinfo{pages}{140--145}.
\newblock


\bibitem[\protect\citeauthoryear{Negara, Codoban, Dig, and Johnson}{Negara
  et~al\mbox{.}}{2014}]%
        {negara2014mining}
\bibfield{author}{\bibinfo{person}{Stas Negara}, \bibinfo{person}{Mihai
  Codoban}, \bibinfo{person}{Danny Dig}, {and} \bibinfo{person}{Ralph~E
  Johnson}.} \bibinfo{year}{2014}\natexlab{}.
\newblock \showarticletitle{Mining fine-grained code changes to detect unknown
  change patterns}. In \bibinfo{booktitle}{\emph{Proceedings of the 36th
  International Conference on Software Engineering}}.
  \bibinfo{pages}{803--813}.
\newblock


\bibitem[\protect\citeauthoryear{Nguyen, Hilton, Codoban, Nguyen, Mast,
  Rademacher, Nguyen, and Dig}{Nguyen et~al\mbox{.}}{2016}]%
        {nguyen2016api}
\bibfield{author}{\bibinfo{person}{Anh~Tuan Nguyen}, \bibinfo{person}{Michael
  Hilton}, \bibinfo{person}{Mihai Codoban}, \bibinfo{person}{Hoan~Anh Nguyen},
  \bibinfo{person}{Lily Mast}, \bibinfo{person}{Eli Rademacher},
  \bibinfo{person}{Tien~N Nguyen}, {and} \bibinfo{person}{Danny Dig}.}
  \bibinfo{year}{2016}\natexlab{}.
\newblock \showarticletitle{API code recommendation using statistical learning
  from fine-grained changes}. In \bibinfo{booktitle}{\emph{Proceedings of the
  2016 24th ACM SIGSOFT International Symposium on Foundations of Software
  Engineering}}. \bibinfo{pages}{511--522}.
\newblock


\bibitem[\protect\citeauthoryear{Nguyen, Nguyen, Nguyen, Nguyen, and
  Rajan}{Nguyen et~al\mbox{.}}{2013}]%
        {nguyen2013study}
\bibfield{author}{\bibinfo{person}{Hoan~Anh Nguyen}, \bibinfo{person}{Anh~Tuan
  Nguyen}, \bibinfo{person}{Tung~Thanh Nguyen}, \bibinfo{person}{Tien~N
  Nguyen}, {and} \bibinfo{person}{Hridesh Rajan}.}
  \bibinfo{year}{2013}\natexlab{}.
\newblock \showarticletitle{A study of repetitiveness of code changes in
  software evolution}. In \bibinfo{booktitle}{\emph{2013 28th IEEE/ACM
  International Conference on Automated Software Engineering (ASE)}}.
  \bibinfo{pages}{180--190}.
\newblock


\bibitem[\protect\citeauthoryear{Nguyen, Nguyen, Dig, Nguyen, Tran, and
  Hilton}{Nguyen et~al\mbox{.}}{2019}]%
        {nguyen2019graph}
\bibfield{author}{\bibinfo{person}{Hoan~Anh Nguyen}, \bibinfo{person}{Tien~N
  Nguyen}, \bibinfo{person}{Danny Dig}, \bibinfo{person}{Son Nguyen},
  \bibinfo{person}{Hieu Tran}, {and} \bibinfo{person}{Michael Hilton}.}
  \bibinfo{year}{2019}\natexlab{}.
\newblock \showarticletitle{Graph-based mining of in-the-wild, fine-grained,
  semantic code change patterns}. In \bibinfo{booktitle}{\emph{2019 IEEE/ACM
  41st International Conference on Software Engineering (ICSE)}}.
  \bibinfo{pages}{819--830}.
\newblock


\bibitem[\protect\citeauthoryear{Osman, Lungu, and Nierstrasz}{Osman
  et~al\mbox{.}}{2014}]%
        {osman2014mining}
\bibfield{author}{\bibinfo{person}{Haidar Osman}, \bibinfo{person}{Mircea
  Lungu}, {and} \bibinfo{person}{Oscar Nierstrasz}.}
  \bibinfo{year}{2014}\natexlab{}.
\newblock \showarticletitle{Mining frequent bug-fix code changes}. In
  \bibinfo{booktitle}{\emph{2014 Software Evolution Week-IEEE Conference on
  Software Maintenance, Reengineering, and Reverse Engineering (CSMR-WCRE)}}.
  \bibinfo{pages}{343--347}.
\newblock


\bibitem[\protect\citeauthoryear{Palomba, Zaidman, Oliveto, and
  De~Lucia}{Palomba et~al\mbox{.}}{2017}]%
        {palomba2017exploratory}
\bibfield{author}{\bibinfo{person}{Fabio Palomba}, \bibinfo{person}{Andy
  Zaidman}, \bibinfo{person}{Rocco Oliveto}, {and} \bibinfo{person}{Andrea
  De~Lucia}.} \bibinfo{year}{2017}\natexlab{}.
\newblock \showarticletitle{An exploratory study on the relationship between
  changes and refactoring}. In \bibinfo{booktitle}{\emph{2017 IEEE/ACM 25th
  International Conference on Program Comprehension (ICPC)}}.
  \bibinfo{pages}{176--185}.
\newblock


\bibitem[\protect\citeauthoryear{Passos, Queiroz, Mukelabai, Berger, Apel,
  Czarnecki, and Padilla}{Passos et~al\mbox{.}}{2018}]%
        {PassosTSE}
\bibfield{author}{\bibinfo{person}{Leonardo Passos}, \bibinfo{person}{Rodrigo
  Queiroz}, \bibinfo{person}{Mukelabai Mukelabai}, \bibinfo{person}{Thorsten
  Berger}, \bibinfo{person}{Sven Apel}, \bibinfo{person}{Krzysztof Czarnecki},
  {and} \bibinfo{person}{Jesus~Alejandro Padilla}.}
  \bibinfo{year}{2018}\natexlab{}.
\newblock \showarticletitle{A Study of Feature Scattering in the Linux Kernel}.
\newblock \bibinfo{journal}{\emph{{IEEE} Transactions on Software Engineering,
  Early Access}} (\bibinfo{year}{2018}), \bibinfo{pages}{1--1}.
\newblock


\bibitem[\protect\citeauthoryear{Rani}{Rani}{2017}]%
        {rani2017detailed}
\bibfield{author}{\bibinfo{person}{Sahil Barjtya Ankur Sharma~Usha Rani}.}
  \bibinfo{year}{2017}\natexlab{}.
\newblock \showarticletitle{A detailed study of Software Development Life Cycle
  (SDLC) models}.
\newblock \bibinfo{journal}{\emph{International Journal Of Engineering And
  Computer Science}} \bibinfo{volume}{6}, \bibinfo{number}{7}
  (\bibinfo{year}{2017}).
\newblock


\bibitem[\protect\citeauthoryear{Ray}{Ray}{2021}]%
        {ray_commit}
\bibfield{author}{\bibinfo{person}{Ray}.} \bibinfo{year}{accessed:
  01.08.2021}\natexlab{}.
\newblock \bibinfo{title}{A commit in Ray}.
\newblock
\newblock
\urldef\tempurl%
\url{https://github.com/ray-project/ray/commit/55fca828ce421b2c11014937a5e2b7d59828a53d}
\showURL{%
\tempurl}


\bibitem[\protect\citeauthoryear{Rodr{\'\i}guez, Haghighatkhah, Lwakatare,
  Teppola, Suomalainen, Eskeli, Karvonen, Kuvaja, Verner, and
  Oivo}{Rodr{\'\i}guez et~al\mbox{.}}{2017}]%
        {rodriguez2017continuous}
\bibfield{author}{\bibinfo{person}{Pilar Rodr{\'\i}guez},
  \bibinfo{person}{Alireza Haghighatkhah}, \bibinfo{person}{Lucy~Ellen
  Lwakatare}, \bibinfo{person}{Susanna Teppola}, \bibinfo{person}{Tanja
  Suomalainen}, \bibinfo{person}{Juho Eskeli}, \bibinfo{person}{Teemu
  Karvonen}, \bibinfo{person}{Pasi Kuvaja}, \bibinfo{person}{June~M Verner},
  {and} \bibinfo{person}{Markku Oivo}.} \bibinfo{year}{2017}\natexlab{}.
\newblock \showarticletitle{Continuous deployment of software intensive
  products and services: A systematic mapping study}.
\newblock \bibinfo{journal}{\emph{Journal of Systems and Software}}
  \bibinfo{volume}{123} (\bibinfo{year}{2017}), \bibinfo{pages}{263--291}.
\newblock


\bibitem[\protect\citeauthoryear{Sharma, Tian, and Lo}{Sharma
  et~al\mbox{.}}{2015}]%
        {sharma2015s}
\bibfield{author}{\bibinfo{person}{Abhishek Sharma}, \bibinfo{person}{Yuan
  Tian}, {and} \bibinfo{person}{David Lo}.} \bibinfo{year}{2015}\natexlab{}.
\newblock \showarticletitle{What's hot in software engineering Twitter space?}.
  In \bibinfo{booktitle}{\emph{2015 IEEE International Conference on Software
  Maintenance and Evolution (ICSME)}}. \bibinfo{pages}{541--545}.
\newblock


\bibitem[\protect\citeauthoryear{Shivaji, Whitehead, Akella, and Kim}{Shivaji
  et~al\mbox{.}}{2012}]%
        {shivaji2012reducing}
\bibfield{author}{\bibinfo{person}{Shivkumar Shivaji}, \bibinfo{person}{E~James
  Whitehead}, \bibinfo{person}{Ram Akella}, {and} \bibinfo{person}{Sunghun
  Kim}.} \bibinfo{year}{2012}\natexlab{}.
\newblock \showarticletitle{Reducing features to improve code change-based bug
  prediction}.
\newblock \bibinfo{journal}{\emph{IEEE Transactions on Software Engineering}}
  \bibinfo{volume}{39}, \bibinfo{number}{4} (\bibinfo{year}{2012}),
  \bibinfo{pages}{552--569}.
\newblock


\bibitem[\protect\citeauthoryear{{\'S}liwerski, Zimmermann, and
  Zeller}{{\'S}liwerski et~al\mbox{.}}{2005}]%
        {sliwerski2005changes}
\bibfield{author}{\bibinfo{person}{Jacek {\'S}liwerski},
  \bibinfo{person}{Thomas Zimmermann}, {and} \bibinfo{person}{Andreas Zeller}.}
  \bibinfo{year}{2005}\natexlab{}.
\newblock \showarticletitle{When do changes induce fixes?}
\newblock \bibinfo{journal}{\emph{ACM sigsoft software engineering notes}}
  \bibinfo{volume}{30}, \bibinfo{number}{4} (\bibinfo{year}{2005}),
  \bibinfo{pages}{1--5}.
\newblock


\bibitem[\protect\citeauthoryear{Spadini, Aniche, and Bacchelli}{Spadini
  et~al\mbox{.}}{2018}]%
        {spadini2018pydriller}
\bibfield{author}{\bibinfo{person}{Davide Spadini},
  \bibinfo{person}{Maur{\'\i}cio Aniche}, {and} \bibinfo{person}{Alberto
  Bacchelli}.} \bibinfo{year}{2018}\natexlab{}.
\newblock \showarticletitle{Pydriller: Python framework for mining software
  repositories}. In \bibinfo{booktitle}{\emph{Proceedings of the 2018 26th ACM
  Joint Meeting on European Software Engineering Conference and Symposium on
  the Foundations of Software Engineering}}. \bibinfo{pages}{908--911}.
\newblock


\bibitem[\protect\citeauthoryear{Ying, Murphy, Ng, and Chu-Carroll}{Ying
  et~al\mbox{.}}{2004}]%
        {ying2004predicting}
\bibfield{author}{\bibinfo{person}{Annie~TT Ying}, \bibinfo{person}{Gail~C
  Murphy}, \bibinfo{person}{Raymond Ng}, {and} \bibinfo{person}{Mark~C
  Chu-Carroll}.} \bibinfo{year}{2004}\natexlab{}.
\newblock \showarticletitle{Predicting source code changes by mining change
  history}.
\newblock \bibinfo{journal}{\emph{IEEE transactions on Software Engineering}}
  \bibinfo{volume}{30}, \bibinfo{number}{9} (\bibinfo{year}{2004}),
  \bibinfo{pages}{574--586}.
\newblock


\bibitem[\protect\citeauthoryear{Zhao, Leung, Yang, Zhou, and Xu}{Zhao
  et~al\mbox{.}}{2017}]%
        {zhao2017towards}
\bibfield{author}{\bibinfo{person}{Yangyang Zhao}, \bibinfo{person}{Hareton
  Leung}, \bibinfo{person}{Yibiao Yang}, \bibinfo{person}{Yuming Zhou}, {and}
  \bibinfo{person}{Baowen Xu}.} \bibinfo{year}{2017}\natexlab{}.
\newblock \showarticletitle{Towards an understanding of change types in bug
  fixing code}.
\newblock \bibinfo{journal}{\emph{Information and software technology}}
  \bibinfo{volume}{86} (\bibinfo{year}{2017}), \bibinfo{pages}{37--53}.
\newblock


\bibitem[\protect\citeauthoryear{Zimmermann, Zeller, Weissgerber, and
  Diehl}{Zimmermann et~al\mbox{.}}{2005}]%
        {zimmermann2005mining}
\bibfield{author}{\bibinfo{person}{Thomas Zimmermann}, \bibinfo{person}{Andreas
  Zeller}, \bibinfo{person}{Peter Weissgerber}, {and} \bibinfo{person}{Stephan
  Diehl}.} \bibinfo{year}{2005}\natexlab{}.
\newblock \showarticletitle{Mining version histories to guide software
  changes}.
\newblock \bibinfo{journal}{\emph{IEEE Transactions on Software Engineering}}
  \bibinfo{volume}{31}, \bibinfo{number}{6} (\bibinfo{year}{2005}),
  \bibinfo{pages}{429--445}.
\newblock


\end{thebibliography}
